\documentclass[12pt,journal,onecolumn,a4paper]{IEEEtran}

\usepackage[ruled,linesnumbered]{algorithm2e}

\SetAlFnt{\small}
\SetAlCapFnt{\small}
\SetAlCapNameFnt{\small}
\SetAlCapHSkip{0pt}
\IncMargin{-\parindent}

\usepackage{enumerate}
\usepackage{subfigure}
\usepackage{booktabs} 
\usepackage{supertabular}

\newtheorem{thm}{Theorem}[section]
\newtheorem{problem}[thm]{Problem}
\newtheorem{definition}[thm]{Definition}
\usepackage[dvipdfmx]{graphicx} 
\usepackage[dvipdfmx]{color}
\usepackage{epstopdf}
\usepackage{epstopdf}
\usepackage{amsmath,amssymb}

\begin{document}


\title{Dynamic Security Analysis of Power Systems\\
by a Sampling-based Algorithm}


\author{Qiang~Wu,~
        T.~John~Koo,~
        Yoshihiko~Susuki
\thanks{Q. Wu and T. J. Koo, Hong Kong Applied Science and Technology Research Institute (ASTRI),
5/F, Photonics Centre, 2 Science Park East Avenue, Hong Kong Science Park, Shatin, Hong Kong; emails: {williamwu,
john.koo}@astri.org; Y. Susuki, Department of Electrical and Information Systems, Osaka Prefecture University, 1-1
Gakuen-cho, Naka-ku, Sakai 599-8531, Japan; email: susuki@eis.osakafu-u.ac.jp.}
\thanks{DOI: doi.org/10.1145/3208093}
}

\maketitle

\begin{abstract}
Dynamic security analysis is an important problem of power systems on ensuring safe operation and stable power supply even when certain faults occur.
No matter such faults are caused by vulnerabilities of system components, physical attacks, or cyber-attacks that are more related to cyber-security, they eventually affect the physical stability of a power system.
Examples of the loss of physical stability include the Northeast blackout of 2003 in North America and the 2015 system-wide blackout in Ukraine.
The nonlinear hybrid nature, that is, nonlinear continuous dynamics integrated with discrete switching, and the high degree of freedom
property of power system dynamics make it challenging to conduct the dynamic security analysis.
In this paper, we use the hybrid automaton model to describe the dynamics of a power system and mainly deal with the index-1 differential-algebraic
equation models regarding the continuous dynamics in different discrete states.
The analysis problem is formulated as a reachability problem of the associated hybrid model.
A sampling-based algorithm is then proposed by integrating modeling and randomized simulation of the hybrid dynamics to search for a feasible execution connecting an initial state of the post-fault system and a target set in the desired operation mode.
The proposed method enables the use of existing power system simulators for the synthesis of discrete switching and control strategies through randomized simulation.
The effectiveness and performance of the proposed approach are demonstrated with an application to the dynamic security analysis of the New England 39-bus benchmark power system exhibiting hybrid dynamics.
In addition to evaluating the dynamic security, the proposed method searches for a feasible strategy to ensure the dynamic security of the system in face of disruptions.
\end{abstract}


\section{Introduction}
Power systems are typical large-scale Cyber-Physical Systems (CPS) \cite{cpsgroup2008,poovendran2010}, especially when referring to the smart grid \cite{nist2009}.
In a power system, the electrical energy produced in generator units is delivered to consumption loads through transmission lines, while the power generation and transmission in the physical world are monitored and controlled by sensing, communication and computation in the cyber-world.
The Supervisory Control And Data Acquisition (SCADA) system has been employed to supervise power systems by collecting data from remote facilities and sending back control commands \cite{esfahani2010}.

\begin{figure}[!b]
\center
\includegraphics[width=0.7\textwidth]{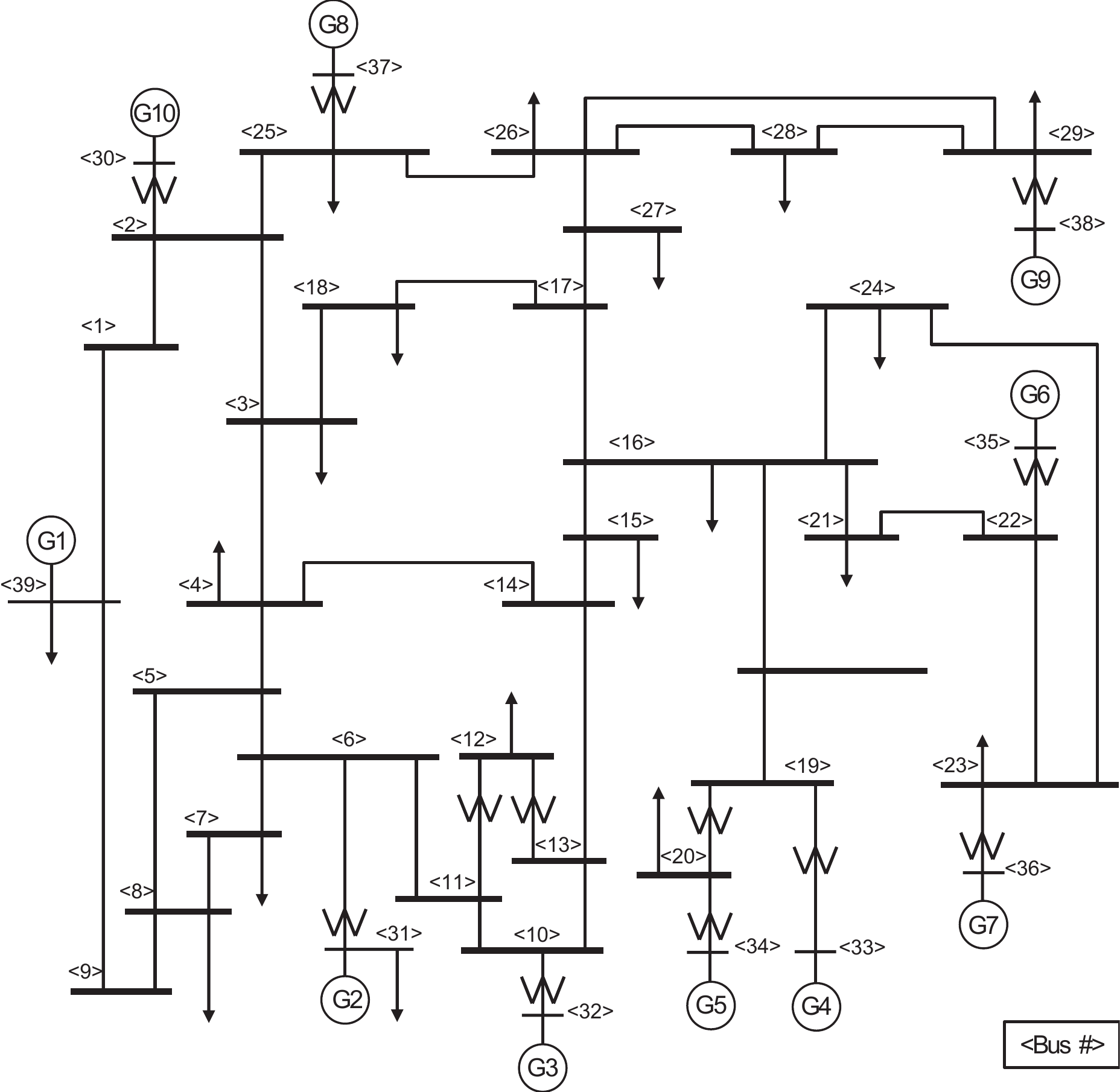}
\caption{One-line diagram of 39-bus New England power system benchmark \cite{athay1979,pai1989}.}\label{f:39bus}
\end{figure}

A power system is by nature continually experiencing disturbances that may be classified as event disturbances and load disturbances according to their effects on the system security \cite{chiang1995}.
Such disturbances may be caused by vulnerabilities of system components, like software defects, or by physical attacks, such as hurricanes, vandalism and terrorist attacks, or by cyber-attacks, including false data injection attacks \cite{yang2014,liang2017}, data integrity attacks \cite{yang2017a,yang2017b} and Denial-of-Service (DoS) attacks \cite{li2017}, etc.
No matter what the reasons are, the disturbances eventually affect the stability and robustness of a power system in the physical world.
Examples of the loss of physical stability include the Northeast blackout of 2003 in North America and the 2015 system-wide blackout in Ukraine.
A software timing defect in the alarm system was recognized as the primary cause of Northeast blackout of 2003, in which more than 508 generating units at 265 power plants were shut down \cite{taskforce2003,poulsen2004,nasa2007}.
The December 2015 Ukraine power grid cyber-attack led to the system-wide blackout in the country, causing 30 substations switched off \cite{kim2016,andy2017}.

Security is one of the priorities of a power system, and it refers to the robustness of the system with respect to imminent faults \cite{fink1978}.
A power system is said to be dynamically secure with respect to a given fault if it is transiently stable \cite{wu1983,kaye1982}, that is, the system can restore its normal operating state following an event disturbance.
Besides the transient stability, different aspects of the system stability, like the short-term voltage stability, are also discussed in the context of dynamic security.

In addition to stability, safety is another term that is always compared with security.
From the view of system engineering, the goal of safety is to guard the physical network and prevent intrusions, while the goal of security is to ensure that the critical functions and the services provided by the network and systems are maintained when the system suffers from disruptions \cite{young2014}.
In power systems, safety is referred in the operation level to ensure the facility safety of power grids and safety of operators,
while security is always used when discussing the ability to provide stable power supply service.

In this paper, we are interested in the Dynamic Security Analysis (DSA) problem of power systems:
\begin{problem}[Dynamic Security Analysis]\label{p:DSA}
The \textit{dynamic security analysis} problem of a power system is to check, when certain faults occur, whether
(1) the system survives the ensuing transient and moves into a steady-state condition; and
(2) neither bus-voltage amplitude nor frequency exceeds its permissible range during transients.
\end{problem}

Instead of the so-called dynamic security assessment in which the fault is taken as a set of certain faults, we focus on analyzing the dynamic responses of the system against one given single fault in the dynamic security analysis problem.
It is straightforward that dynamic security analysis serves as a basis for dynamic security assessment.
In parallel to checking the system security, we also search for a feasible solution including operation modes switching strategies in the cyber-world and control input selection methods in the physical world for the system to resume automatically after a fault occurs.
In this study, the consequence of the fault on the system is emphasized rather than the cause of the fault.

Existing methods for the dynamic security analysis may be roughly classified into direct methods and time-domain approaches.
Direct methods, including the techniques of linearization and bounding \cite{kaye1982}, Liapunov functions \cite{pavella1977}, transient energy functions \cite{athay1979}, transient energy margins \cite{fouad1981}, transient security indices \cite{pavella1982}, BCU (Boundary of stability region based Controlling Unstable equilibrium point) \cite{chiang1995,chiang2010}, EEAC (Extended Equal-Area Criterion) \cite{xue1989}, are fast in computation and provide the degree of stability or instability, together with useful information regarding the preventive control.
But they may be not applicable to those models for which it is hard to explicitly derive energy functions, and could not provide time responses of the post-fault system.
In contrast, time-domain approaches are applicable to general system models and provide detailed system dynamic responses through Step-By-Step (SBS) simulation.
But they may not offer the strategy for the preventive control.
In addition, efforts based on trajectory sensitivity \cite{hiskens2000}, pattern recognition method \cite{saito1975}, machine learning \cite{chu2004,shenoy2016} have been contributed in this field.

The one-line diagram in Fig. \ref{f:39bus} shows a typical benchmark of power systems, the 39-bus New England system \cite{athay1979,pai1989},
where synchronous generators and loads are connected to buses linked via high-voltage lines.
The continuous-time dynamics of power systems are normally governed by a nonlinear Differential-Algebraic Equation (DAE), since the swing of rotors in generators is represented by nonlinear differential equations, while the power balance of loads and supplies on buses is described by algebraic equations.
On the other hand, different connection status of components result in different operation modes of the system, and the system is governed by
different continuous dynamics in accordance with the operation mode.
The complicated dynamics of power systems exhibit both hybrid and nonlinear nature and have high degree of freedom, making the dynamic security analysis  a challenging problem.

One key enabler for dynamic security analysis lies in detailed models of target power systems.
Large-scale power systems consist of many interconnected components and complex networks, and thus require special efforts to build the system models.
Considering its hybrid nature, several researchers proposed to adopt mixed logical dynamical systems \cite{bemporad1999} and the theory of hybrid systems \cite{henzinger1996,tomlin2000,lygeros2003,lygeros2006} in modeling, analysis and control of power systems, see Ref. \cite{fourlas2004,hikihara2005,susuki2009,susuki2012} and references therein.
Through proper modeling of the system dynamics, the dynamic security analysis problem could be formulated and addressed as a reachability problem of the associated hybrid model \cite{susuki2015,wu2016}.

Sampling-based algorithms, such as Probabilistic RoadMap (PRM) \cite{amato1996,kavraki1996} and Rapidly-exploring Random Tree (RRT) \cite{lavalle1998,lavalle2006,lavalle2001}, provide a practical computational approach to solving robotic path and motion planning problems with many degrees of freedom.
The idea is probability-based, so that it overcomes the computational limitation caused by high dimension and nonlinearity of the target dynamics.
In particular, RRT provides an efficient data structure and sampling scheme for high-dimensional problems with differential constraints and complicated constraints \cite{lavalle1998,lavalle2000,lavalle1999,lavalle2001a}.
Although, initially proposed for continuous problems, RRT has been extended to handle control and verification problems with hybrid dynamics \cite{branicky2003,bhatia2004,dang2006,dang2009,plaku2007,plaku2009,xu2010}.
In Ref. \cite{susuki2015,wu2016}, the authors introduced the RRT algorithm to the field of power systems and developed a computational approach to the dynamic security analysis problems.

In this paper, we use the hybrid automaton model to describe the hybrid dynamics of power systems, and mainly deal with DAE models that could be equivalently transformed into an Ordinary Differential Equation (ODE) with once differentiation, i.e., the so-called index-1 DAE \cite{brenan1996},
regarding the continuous dynamics.
As an alternative to direct mathematical modeling, we propose to model the continuous dynamics in each discrete state with existing tools and then embed them in the automaton for discrete switching to build the whole hybrid model.
The dynamic security analysis problem is formulated as searching for a feasible execution of the hybrid model of a target system.
The system is proved to be dynamically secure if there exists such an execution that starts from the current physical state and operation mode
and arrives at a target set in the desired operation mode eventually.
By integrating power system simulators, a sampling-based computational approach is developed to conduct the search of discrete switching and control strategies in the high-dimensional hybrid state space.

The main contribution of this work lies in that the proposed modeling and computation method enables existing simulators for the formal verification and controller synthesis through the randomized simulation.
Take the field of power systems as an example.
A large set of simulators validated in realistic problems, like Power System Toolbox (PST) \cite{chow1992} and Power System Analysis Toolbox (PSAT) \cite{milano2005}, have been used for the assessment of system dynamic performance.
All these simulators conduct time-domain simulation to verify the control strategy defined in advance by simulating the time responses, but neither of them was directly applied to check the correctness of power systems in the sense of formal method or to synthesize the correct behavior by control.
In this paper, by using the proposed method with PSAT, we demonstrate for the first time that the dynamic security analysis with practical simulators is possible.
Besides, since the developed approach is module-based, the simulator could be replaced with any other one of interest, implying the potential of the approach to practical and large-scale problems.

The rest of this paper is organized as follows.
Section \ref{sec:state.of.art} discusses some related works in the filed of DSA and sampling-based algorithms.
In Sec. \ref{sec:modeling}, the DAE and hybrid automaton models are introduced for the continuous and hybrid dynamics of power systems, respectively.
A sampling-based algorithm for the hybrid dynamics is developed in Sec. \ref{sec:sampling.based.algorithms}
with an investigation on its properties of coverage and computational complexity.
Section \ref{sec:application} shows a demonstration of the proposed algorithm
on the 39-bus New England System, as shown in Fig. \ref{f:39bus}, exhibiting hybrid dynamics.
The conclusion of this paper with a brief summary and future work is drawn in Sec. \ref{sec:conclusion}.

\section{Related Works}
\label{sec:state.of.art}
The threat model of power systems has been studied in the context of CPS security under a unified framework \cite{abdu2017}, which consists the threats, vulnerabilities, attacks and controls from the security perspective and the cyber-, physical and cyber-physical components from the CPS components perspective.
Cyber-security of power systems \cite{li2017} plays a significant role in managing power grid operations, due to the integration of information and communication technologies in power systems.
The mechanism of typical cyber-attacks, such as false data injection attacks, data integrity attacks and DoS attacks, etc., has been studied using different models for power systems, together with detection and prevention strategies.
In addition, testbeds \cite{cintuglu2017} have been established to evaluate vulnerabilities of smart grids.

Generally, state-of-the-art DSA tools rely heavily on deterministic computation methods \cite{morison2004}, such as the aforementioned EEAC, BCU, SBS, etc.
Security limits are searched based on exhaustively examining many predefined contingencies using a rigorous approach.
Advantages of such methods are the accurate results available at the expense of computation time, since precise models of the system dynamics are used as the basis.
With the development of data science, the probabilistic methods, including machine learning and artificial intelligence \cite{shenoy2016,chu2004}, are recently proposed again to DSA, utilizing accumulated knowledge and data.
Some probabilistic models for the causal relation could be trained using certain learning methods based on data obtained via simulation, and the built models are further employed to conduct the on-line DSA.
The advantage of such methods are the rapid results, while the reliability of the obtained models is still in study, and simulation still plays a fundamental role on supplying the data.
In addition, a rigorous and concrete strategy to maintain the system security is not provided in the state-of-the-art tools and methods.

In the field of robotic path and motion planning, there have been a large number of variants of RRT, like RRT-connect \cite{kuffner2000}, RRT$^*$ \cite{karaman2010}, Linear Quadratic Regulation based RRT$^*$ (LQR-RRT$^*$) \cite{perez2012}, etc., and implemented libraries on sampling-based algorithms, like Open Motion Planning Library (OMPL) \cite{sucan2012}.
RRT-connect improves the efficiency of RRT by incrementally building searching trees rooted at the start and the goal configurations.
RRT$^*$ generates an asymptotically optimal trajectory by rewiring the tree as it discovers new lower-cost paths reaching the nodes that are already in the tree.
LQR-RRT$^*$ finds optimal plans in domains with complex or underactuated dynamics by locally linearizing the domain dynamics and applying linear quadratic regulation.
However, neither of them is directly applicable to the power system application, due to the large scale and nonlinearity of power system dynamics.

To the best of our knowledge, the application of the RRT idea to dynamic security analysis of power systems is firstly reported in Ref. \cite{susuki2015} and then strengthened in Ref. \cite{wu2016}.
The analysis problem was formulated as checking the reachability of a mathematical model representing dynamic performances of a target power system.
The RRT algorithm was used as a computational approach to searching for a feasible trajectory connecting an initial state possibly at a lower security level and a target set with a desirable higher security level.
Case studies on the Single Machine-Infinite Bus (SMIB) system and the two-area system for frequency control problems demonstrated the performance and effectiveness of the proposed approach.

In the previous work, the system dynamics were modeled in a explicit manner, that is, the ODEs for each component were exhaustively written down
and the whole system model was then integrated manually for the continuous dynamics based on field knowledge.
Since a new system model is required once the connection status of the system network changes or a component is replaced, the application of such explicit method is limited by the modeling step.
In this paper, through the integration of the original RRT and an existing power system simulator, we enhance the previous work on DSA and make it feasible to apply the RRT idea to larger and practical-scale power systems.

\section{Modeling}
\label{sec:modeling}
In this section, we first introduce the DAE model of power system dynamics, and then discuss the modeling and simulation of the dynamic responses with existing tools.
Following that, the mathematical model of hybrid automaton is introduced to describe the hybrid dynamics in Sec. \ref{sec:hybrid.automaton}.
At the end of this section, the problem of dynamic security analysis of a power system is formulated as checking of the reachability of the hybrid automaton model representing the hybrid dynamics.

\subsection{Power System Modeling}
The dynamic responses of a power system can be described by a system of equations \cite{sauer1998}:
\begin{eqnarray}\label{e:DAE.general}
F (\dot x, x, u) = 0,
\end{eqnarray}
where $F: \mathbb{R}^n \times \mathbb{R}^n \times \mathbb{R}^m  \to \mathbb{R}^n$ is a vector-valued function,
$x \in \mathbb{R}^n$ and $u \in \mathbb{R}^m$ represent the variables and controls, respectively, $m, n \in \mathbb{N}_+$,
and $\dot x$ denotes the derivative of $x$ with respect to the time $t$, i.e., $\dot x =  {\rm d} x / {\rm d} t $.
If the Jacobian matrix $ \partial F / \partial \dot x $ is nonsingular, then an explicit ODE system $\dot x = f (x,u)$ could be obtained according to the implicit function theorem.
Otherwise, the system of equations  $F ( \dot x,x,u) = 0$ defines a DAE.

Normally, the DAE for a power system can be written in the semi-explicit form \cite{sauer1998} (also called ODE with constraints):
\begin{eqnarray}\label{e:DAE}
 \frac{{\rm d} y}{ {\rm d} t} = \phi(y,z,u), \quad\quad 0 = \psi(y,z,u),
\end{eqnarray}
where
$y$ represents the differential variables including rotor angles $\delta$, rotor speeds $\omega$ of generators, and so on,
$z $ the algebraic variables, typically including bus-voltage phases $\theta$ and amplitudes $v$,  etc.,
$\phi $ the differential equations representing the electro-mechanical behaviors of synchronous generators and controllers,
and $\psi $ the algebraic equations governing the active and reactive power balance at buses.
The variables $y$ and $z$ together constitute $x$ in Eq. (\ref{e:DAE.general}), i.e., $(y,z) =x$.

The DAE in the form of Eq. (\ref{e:DAE}) can be transformed into an explicit ODE if the derivatives of the algebraic variables, $\dot z$, can be uniquely determined.
The condition that the Jacobian matrix $\partial \psi / \partial z$ is nonsingular serves as a sufficient condition which enables the transformation.
The minimum number of times of differentiation to the system that is required to solve for $\dot x$ uniquely in terms of $x$ is called \emph{index} of the DAE \cite{brenan1996}, and the index measures the distance from a DAE to its related ODE.
Notice Eq. (\ref{e:DAE}) is an index-1 DAE if $\det \partial \psi / \partial z \neq 0$, where $\det$ is the determinant of a matrix.

In this paper, we mainly concern with the ODE and the class of DAE which is equivalent to an ODE after some transformations.
Hence, we assume either the matrix $ \partial F / \partial {\dot x} $ in Eq. (\ref{e:DAE.general}) or the matrix $\partial \psi / \partial z$ in Eq. (\ref{e:DAE}) is nonsingular.
Since in both cases the system $F (\dot x, x, u) = 0$ is equivalent to an ODE $\dot x = f (x,u)$, we would define the hybrid automaton and its semantics based on the ODE form as the continuous dynamics in the following of this paper.
Besides, we assume the ODE to be \textit{Lipschitz} continuous, such that, the existence and uniqueness of solution of the initial value problem,
$\dot x= f(  x, u),~x(0) = x_0$, can be guaranteed.

There have been a large set of simulators in the field of power systems, like PST and PSAT \cite{milano2005}, for the modeling and analysis of
the system behaviors.
PSAT is a MATLAB-based open source toolbox that has been commonly accepted.
Typical physical components, such as generators, buses, lines and so on, are provided with built-in static and dynamic models in PSAT,
enabling users to build their system models with these given components and system networks.
In PSAT, dynamic behaviors of systems are modeled as a set of nonlinear DAE as Eq. (\ref{e:DAE}).
These equations are implicitly embedded in the PSAT model, while values of variables, equations and the Jacobian matrices are recorded in a special structure named DAE.
Further analysis of the system can be conducted based on the DAE structure via static analysis and time domain simulation.

In a power system, each discrete state corresponds to a particular connection status of the system network, while the status is characterized with some parameters of components.
The switching between discrete states results in the hybrid dynamics of a power system.
In order to capture the hybrid nature, the system model with all its components well connected is built through configuring the system network and parameters.
The model representing the continuous dynamics is then constructed in its DAE structure automatically with power flow computation in PSAT, together with other structures for system components.
These structures are recorded as the basic model to construct the whole hybrid model by integrating into an automaton for discrete states and jumps.
Whenever jumping to a new discrete state, the system reaches new (continuous and discrete) states following the automaton.
The basic model is updated according to the parameters characterizing the new states and control inputs for the corresponding continuous dynamics model.
The hybrid model for a power system can be constructed correspondingly.
We next define the mathematical model of hybrid automaton rigorously for the hybrid dynamics.

\subsection{Hybrid Automaton}
\label{sec:hybrid.automaton}
A system with hybrid dynamics can be formally modeled using a mathematical model called a hybrid automaton \cite{henzinger1996,tomlin2000,lygeros2003}.
A hybrid automaton can be defined as follows:

\begin{definition}[Hybrid Automaton]\label{def:hybrid.automaton}
A \emph{hybrid automaton} $\mathbb{H}$ is a collection
$\mathbb{H}=(Q,X,U,f,Init,E,G,R)$, where
$Q = \{ q_1, q_2, \dots \}$ is a set of \emph{discrete states};
$X \subseteq \mathbb{R}^n$ is a set of \emph{continuous states};
$U$ is a set of \emph{inputs};
$f (\cdot,  \cdot, \cdot) : Q \times   X \times U \to \mathbb{R}^n$ is a \emph{vector field};
$Init \subseteq  Q \times X $ is a set of \emph{initial states};
$E \subseteq Q \times Q$ is a set of \emph{edges};
$G (\cdot): E \to 2^X$ is a \emph{guard condition};
$R (\cdot, \cdot, \cdot) : E \times X  \to 2^X $ is a \emph{reset map}.
\end{definition}

In the definition, $2^X$ denotes the power set, namely the set of all subsets, of $X$.
The pair $(q,x)$, where $q \in Q$ and $x \in X$, is called the \emph{hybrid state} and denoted by $s$, while $S:= Q \times X$ is the \emph{hybrid state space} of $\mathbb{H}$.
We assume that both the number of the discrete states and the dimension of continuous states are finite.
With abuse of notations, the vector field in the hybrid automaton is denoted by $f$, as in the previous section, but with an additional variable $q \in Q$ to indicate the discrete state.
These definitions are applicable to the hybrid automaton with the continuous dynamics in the DAE form.

A \emph{hybrid time trajectory} $\tau:=\{I_i\}^N_{i=0}$ is a finite or infinite sequence of intervals on the real line, such that:
for all $i<N$, $I_i=[\tau_i,\tau'_i]$;
if $N<\infty$, $I_N=[\tau_N,\tau'_N]$ or $I_N=[\tau_N,\tau'_N)$;
for all $i$, $\tau_i \leq \tau'_i$ and $ \tau'_i =\tau_{i+1}$.

An \emph{execution} of the hybrid automaton $\mathbb{H}$ is a hybrid \emph{trajectory} $\chi := (\tau, q, x, u)$ with
initial condition: $( q(\tau_0),x(\tau_0) )\in Init$;
continuous evolution: for all $i$ with $\tau_i < \tau'_{i}$,
  $q(\cdot) \in Q$ is constant, $u(\cdot) \in U$ is piecewise continuous,
  $x(\cdot) \in X$ is a solution to the ODE $\dot x  = f(q,x,u)$ over $[\tau_i, \tau'_{i}]$;
discrete evolution: for all $i < N$, $(q(\tau'_{i}), q(\tau_{i+1})) \in E$, $x (\tau'_{i}) \in G ( q(\tau'_{i}), q(\tau_{i+1}) )$, and
  $ x(\tau_{i+1}) \in R (  q(\tau'_{i}), q(\tau_{i+1}), x(\tau'_{i})  )$.

A hybrid automaton $\mathbb{H}$ accepts an execution $\chi$ if $\chi$ satisfies the above conditions.
The semantics of a hybrid automaton $\mathbb{H}$ is characterized by the executions it accepts.
In this paper, we are interested in the finite execution $\chi$ in which the hybrid time trajectory $\tau$ is a finite sequence ending with a closed interval.
More detailed notions and properties of hybrid automaton $\mathbb{H}$ can be found in Ref. \cite{henzinger1996,tomlin2000,lygeros2003} and references therein.

Given a finite hybrid time trajectory $\tau $, let $u$ be a {\em control input sequence} along the trajectory $\tau$, such that:
on each interval $[\tau_i, \tau'_i]$ where $\tau_i < \tau'_i$, $u (\cdot) \in U$ is a piecewise continuous function and applied to the continuous variables;
for each $i$ with $\tau_i = \tau'_i$, $u$ is a discrete input that may result in a transition of discrete state, say, $q \to q'$,
and the input is be denoted by the final state after the transition, i.e., $u (\tau_i) := q'$.

Note that with the abuse of notations, both the continuous inputs and input sequence are denoted by $u$ here.
We denote an execution starting at $( q(\tau_0),x(\tau_0) )$ and following the control input sequence $u$ along the hybrid time trajectory $\tau$ by $\chi ( \tau, q(\tau_0),x(\tau_0), u )$ in this paper, and further denote its first and last states by $\chi_{\tau_0} ( \tau, q(\tau_0),x(\tau_0), u ) $ and $\chi_{\tau'_N} ( \tau, q(\tau_0),x(\tau_0), u )$, respectively.
Namely, $\chi_{\tau_0}  ( \tau, q(\tau_0),x(\tau_0), u ) := ( q(\tau_0), x(\tau_0))$ and $\chi_{\tau'_N} ( \tau, q(\tau_0),x(\tau_0), u ) := (q(\tau'_N), x(\tau'_N))$.
These notations of the first and last states are introduced for the hybrid RRT algorithm in Sec. \ref{sec:sampling.based.algorithms}.

The objective of dynamic security analysis is to investigate the robustness of the system in its current operation mode by checking if the system can attain its normal operation status with all constraints and control objectives fulfilled by taking proper control actions.
Let the hybrid dynamics of a power system be modeled with a hybrid automaton $\mathbb{H}$.
Given an initial state $s_{\rm init}$ describing the current physical state and operation mode of the system, and a target set $S_{\rm goal}$ the desired physical states and operation mode, the dynamic security analysis problem is equivalent to determining the reachability of the hybrid automaton in the sense:
whether there exists a pair of hybrid time trajectory $\tau = \{ I_i \}_{i=1}^N$ and hybrid execution $\chi ( \tau, q, x, u)$ of $\mathbb{H}$, s.t.
$
\exists t \in I_i \textrm{ for some } i  \leq  N,  (q(\tau_0),x(\tau_0)) = s_{\rm init} \textrm{ and } (q(t),x(t)) \in S_{\rm goal}.
$

In the automaton, unsafe regions are described as obstacles, while the (free) hybrid state space modeled as their complement.
Hence, only the feasible executions in the state space are taken into account in analyzing the dynamic security.

\section{Sampling-based Algorithms}
\label{sec:sampling.based.algorithms}
Sampling-based algorithms, especially RRT \cite{lavalle1998,lavalle2001}, provide a practical computational approach to the path and motion planning problems with high degrees of freedom and nonlinear dynamics.
In this paper, we focus on the RRT algorithm for hybrid automaton and propose to utilize existing modeling tools and simulators in the RRT computation for large-scale problems.
This section first introduces the RRT algorithm for hybrid models with simulators and then discusses the properties of probabilistic completeness and time complexity briefly.

\subsection{RRT Algorithm}
\label{sec:rrt.algorithm}
Through iteratively applying the control input that drives the system to evolve towards a randomly selected sample in the search space, a RRT searching tree can be incrementally expanded from a given initial state with the construction algorithm shown in Algorithm \ref{alg:rrt}.

\IncMargin{1em}
\begin{algorithm}[tbh]
\SetAlgoNoLine

\Indm
\KwIn{$s_{\rm init}$, $K$, $\Delta t$;}
\KwOut{$\mathcal{T}$;}

\Indp
  \BlankLine
 $\mathcal{T}.\textrm{init}(s_{\rm init})$\;
 \For{
 $k = 1 \to K$
 }
 {
 $s_{\rm rand} \gets \textrm{RandomState}()$\;
 $s_{\rm near} \gets \textrm{NearestNeighbor}(s_{\rm rand},\mathcal{T})$\;
 $(s_{\rm new},u_{\rm new}) \gets \textrm{NearestNewNeighbor}(s_{\rm near}, s_{\rm rand}, U, \Delta t)$\;
 $\mathcal{T}.\textrm{AddVertex}(s_{\rm new})$\;
 $\mathcal{T}.\textrm{AddEdge}( s_{\rm near},s_{\rm new},u_{\rm new})$\;
 }
 \Return $\mathcal{T}$

\Indp
\caption{RRT Algorithm}
\label{alg:rrt}
\end{algorithm}
\DecMargin{1em}

In the algorithm, $s_{\rm init}$ is the initial state,
$s_{\rm rand}$ a random state,
$s_{\rm near}$ the nearest vertex to the random state,
$s_{\rm new}$ the new vertex,
$U$ the set of continuous input,
$u_{\rm new}$ the control input corresponding to $s_{\rm new}$,
and $\mathcal{T}$ the RRT searching tree.
The parameters $K$ and $\Delta t$ are the number of iterations and the sampling interval of time, respectively.
Let the search space $S$ be equipped with a distance function $\rho: S \times S \to \mathbb{R}$, then the RRT searching tree $\mathcal{T}$ with root $s_{\rm init}$  follows the key procedures:
\begin{enumerate}[{\em Step 1.}]
\item Initialize the searching tree $\mathcal{T}$ with the initial state $s_{\rm init}$ (line 1).

\item In the beginning of each loop, select a random sample $s_{\rm rand}$ with uniform sampling in the search space $S$ (line 3).

\item Take the vertex on the tree $\mathcal{T}$ with the shortest distance to the random sample $s_{\rm rand}$ as the nearest node $s_{\rm near}$ (line 4):
        $$
        s_{\rm near} :=  \mathop{\arg \min}_{v \in \mathcal{T}}   \rho(s_{\rm rand}, v).
        $$

\item Let the system evolve from $s_{\rm near}$ and according to its dynamic model to obtain the set of potential new states that the system could arrive at $\Delta t$, denoted as $S_{\rm P} (s_{\rm near} )$.
        Choose the nearest one to the random sample $s_{\rm rand}$ in this set as the new state $s_{\rm new}$, and denote the corresponding control input as $u_{\rm new}$ (line 5):
        $$
        s_{\rm new} :=   \mathop{\arg \min}_{s \in S_{\rm P} (s_{\rm near})}  \rho (s_{\rm rand}, s).
        $$
        With regards to applying the controls, if the input set $U$ is finite, all its elements are exhaustively used.
        Otherwise, if the set $U$ is an infinite one, special methods, such as sampling, can be employed to take finite samples in the computation.
        Note that a control $u \in U$ may be either a constant or a continuous function over the time interval $[0, \Delta t]$.
        Unsafe regions and obstacles are described as constraints in the state space.
        Collision detection is conducted in this step, such that only new states with feasible executions are taken into account.

\item Add the state $s_{\rm new}$ to the tree $\mathcal{T}$ as a new vertex. Record the edge $(s_{\rm near}, s_{\rm new})$ in $\mathcal{T}$, together with the input $u_{\rm new}$ (lines 6-7).
\end{enumerate}

As we are dealing with hybrid dynamics in high-dimensional hybrid state space, we make the following extensions on the sampling strategy and the distance function, in addition to the computation framework.
\begin{enumerate}
\item \textit{Sampling strategy.}
Different methods, such as the goal-bias sampling \cite{lavalle2001}, can be employed for specified purposes.
In this paper, random samples are selected in the hybrid state space $S$ with uniform sampling for both discrete and continuous state variables.

\item \textit{Distance function.}
The distance between two hybrid states is normally defined as the weighted sum of distances between the two discrete states and between the continuous states \cite{branicky2003}.
Since the switching of discrete state neither changes the continuous state nor costs time in the power system application, we set the distance between discrete states as zero, and define the distance function on the nonlinear continuous state space in Sec. \ref{sec:application}.

\item \textit{Computation framework.}
We adopt the classical RRT computation framework but make some modification on the extension step (Step 4), in addition to utilizing the hybrid sampling strategy and the hybrid distance function in the computation.
As for the hybrid system, not only the states that the system could arrive at after a continuous evolution for $[0, \Delta t]$, but also the ones
that the system arrives at after a discrete jump and the continuous evolution
are taken into account.

In the computation, the potential children nodes of $s_{\rm near}$ are selected as all the last states of the system executions starting from
$s_{\rm near}$ following all possible control input sequences along the hybrid time trajectory $\{ 0 , [0, \Delta t]\}$, that is,
\begin{eqnarray*}
S_{\rm P} (s_{\rm near}) = \{ s \in S :  s := \chi_{\Delta t} ( \{ 0, [0, \Delta t] \} , s_{\rm near}, \{q, u\}), \textrm{~for~} q \in Q \textrm{~and~} u \in U \}.
\end{eqnarray*}

\end{enumerate}

\begin{figure}[tbh]
\center
\includegraphics[width=0.6\textwidth]{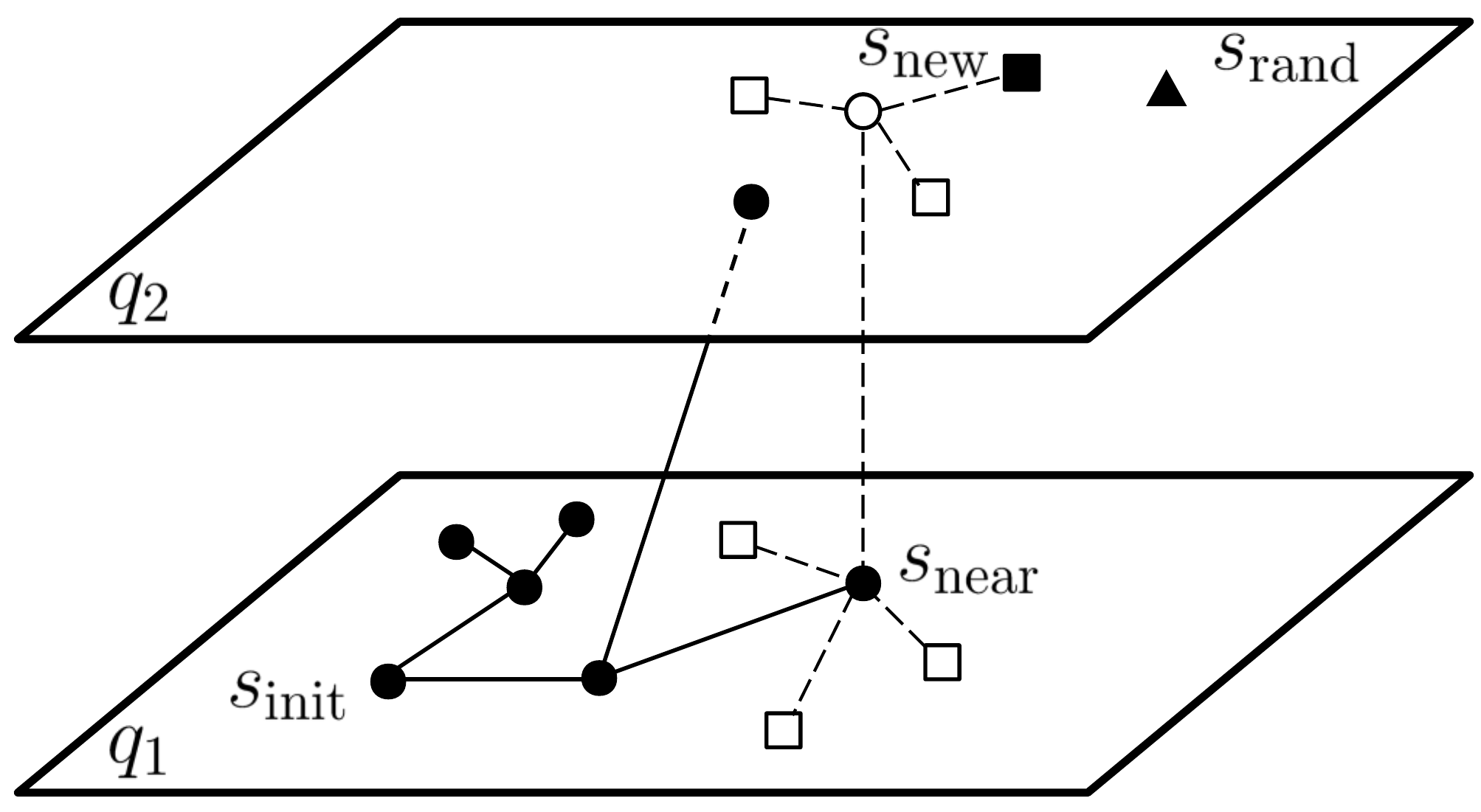}
\caption{Illustration of the new state extension of RRT in hybrid state space.} \label{f:rrt.hybrid}
\end{figure}
Figure \ref{f:rrt.hybrid} illustrates the RRT algorithm for a hybrid automaton with two discrete states, namely $q_1$ and $q_2$.
In this figure, the dark dots are the vertices that have been on the tree already, and the triangle shows a random state with the discrete state $q_2$.
The squares are the potential new states from $s_{\rm near}$, i.e., the set $S_{\rm P} (s_{\rm near})$.
The solid square shows the new state $s_{\rm new}$  selected from $S_{\rm P}(s_{\rm near})$.

Regarding large-scale systems like the power system we deal with in this paper, we propose to utilize existing modeling tools to model the continuous dynamics of the system, as discussed in Sec. \ref{sec:modeling}.
Since the DAE of the system dynamics is implicitly embedded in the model, we adopt simulators for the randomized simulation and propose an integrated computation framework with modeling tools and simulators.

In each simulation with the implicit model, the modeling tool is initialized with a basic model of the system constructed in advance for the continuous dynamics.
Parameters and control inputs are then passed to the tool.
The parameters include two parts.
One part is for configuration of the tool,  like the type of numerical solver and the integration step size in computation, while the other part determines the discrete state of the system, like the connection status for a specified component, and the control input.
Following that, the ODE (or DAE) model describing continuous dynamics of the system is updated with these parameters and also the nearest node $s_{\rm near}$ to realize the potential reset map.
Time-domain simulation is then conducted in the simulator on the predefined time interval $[0,\Delta t]$ to obtain the potential new state with the given pair of discrete state and control input $(q_i, u_j)$, for $q_i \in Q$ and $u_j \in U$.

In addition, since time-domain simulation consumes much more time than other procedures in the whole computation, all the simulation results started from a vertex are saved at the first time when the vertex is selected as the nearest node $s_{\rm near}$, including the finial states and other necessary structures in the model.
Let a vertex $s_j \in \mathcal{T}$ be the nearest node at the $k$-th loop, i.e., $s_{\rm near} := s_j$.
Assume this is the first time $s_j$ is selected as $s_{\rm near}$.
The set of results for $s_j$ is recorded as $S_{\rm P}(s_j)$, while the state $s_j$ recorded in a set $S_{\rm N}$.
If the vertex $s_j$ is selected as the nearest node again at the $k'$-th loop for some $k' >k$, these corresponding results $S_{\rm P}(s_j)$ are reused directly to expand the new vertex without repetition of simulation, in order to save the computation time.

\subsection{Probabilistic Completeness}
One significant advantage of RRT algorithm over other planning algorithms is the so-called probabilistic completeness \cite{lavalle2001}.
It ensures the RRT searching tree converges, in the sense of probability, to a uniform coverage of the free configuration space, that is, the complement of obstacles in search space.
However, this property is based on the assumption that the system could reach any point in the search space from the initial point.
This assumption does not always hold, particularly, for many nonlinear systems.
Actually, this is exactly the problem to be verified with reachability analysis.

As an alternative, a property called reachability completeness is proposed in Ref. \cite{dang2006,dang2009}.
This property tells that the RRT searching tree converges to a uniform coverage of the reachable set instead of the whole state space of the system.
The reachablility completeness property indicates that for any reachable state $s$ of the system and any real number $\varepsilon > 0$, the probability of the RRT tree $\mathcal{T}$ containing a vertex $v$ whose distance to $s$ is smaller than $\varepsilon$ converges to one as the number $K$ goes to infinity:
$$
\forall s \in Reach, \forall \varepsilon > 0,  \lim_{K \to \infty} \mathsf{P} [~\exists v \in \mathcal{T}^K, \textrm{  s.t. } \rho(s,v) \leq \varepsilon] = 1 ,
$$
where $\mathsf{P}$ is the probability, $\mathcal{T}^K$ is an RRT tree with $K$ vertices, $\rho $ is the distance function, $Reach \subseteq S$ is the reachable set, which is the set of states that the system starting from the initial state could arrive within a time interval.

It is proved in Ref. \cite{dang2006} that the reachability completeness is validated with two sufficient conditions:
\begin{enumerate}
\item the probability of each vertex on $\mathcal{T}$ being selected as $s_{\rm near}$ in each loop is non-null;
\item the probability of each reachable direction being selected in each loop is non-null.
\end{enumerate}
The first condition is fulfilled with the uniform sampling method in the sampling procedure (in Step 2 of RRT computation), and the second one is also satisfied if the input set $U$ is a finite set and the probability of each $u \in U$ being selected is greater than zero (in Step 4 of RRT computation).

The reachability completeness is a property of the RRT computation regarding less the nonlinearity and scale of the system dynamics.
For a given system, once its initial condition and the control input set are fixed, the reachable set is indeed determined.
If the two conditions are fulfilled, then the RRT computation generates a (probabilistically complete) coverage of the reachable set.
Therefore, it is possible to adopt the RRT computation in the reachability analysis, particularly, for nonlinear and large-scale problems,
taking advantage of the reachability completeness property and computational efficiency.

\subsection{Time Complexity Analysis}\label{sec:time.complexity}
Disregarding the time of initialization, the computation time to construct a RRT searching tree within $K$ iterations is the sum of time cost in each procedure:
\begin{eqnarray*}
T(K)= \sum_{k=1}^K t (k)  = \sum_{k=1}^K  \big( t_{\rm Step2} (k) + t_{\rm Step3} (k) + t_{\rm Step4} (k) + t_{\rm Step5} (k) \big) \text{,}
\end{eqnarray*}
where $t_{\rm Step2} (k)$, $ t_{\rm Step3} (k)$, $t_{\rm Step4} (k)$, $t_{\rm Step5} (k)$ correspond to the time costs of the sampling (Step 2 of RRT computation), finding the nearest vertex (Step 3), extending a new vertex (Step 4) and adding the new vertex procedures (Step 5) in the $k$-th loop, respectively.
In each loop, $t_{\rm Step2} (k)$, $t_{\rm Step4} (k)$ and $t_{\rm Step5} (k)$ are independent of $k$ and bounded with some positive constants $c_s$, $c_e$ and $c_a$, respectively.
In finding the nearest neighbor, all existing vertices are used to calculate and compare the distances from a new sample, so  $ t_{\rm Step3} (k) \leq k c_n$, for some constant $c_n>0$.
Then, the total computation time can be estimated as follows:
\begin{eqnarray*}
T(K) & \leq & \sum_{k=1}^K   ( c_s + c_a + c_e + k c_n  ) \\
& = & ( c_s + c_a + c_e   ) K + \frac{c_n}{2} (K^2 - K) \\
& = & \frac{c_n}{2} K^2 + ( c_s + c_a + c_e -  \frac{c_n}{2} ) K \text{.}
\end{eqnarray*}
That is, the time complexity of this algorithm is $O(K^2)$ in the worst case scenario.

\begin{figure}[htb]
\centering
\subfigure[\text{$K$ from 100 to 1000.} ]{
\label{f:rrt.time.complexity.c1}
\includegraphics[width=0.45\textwidth]{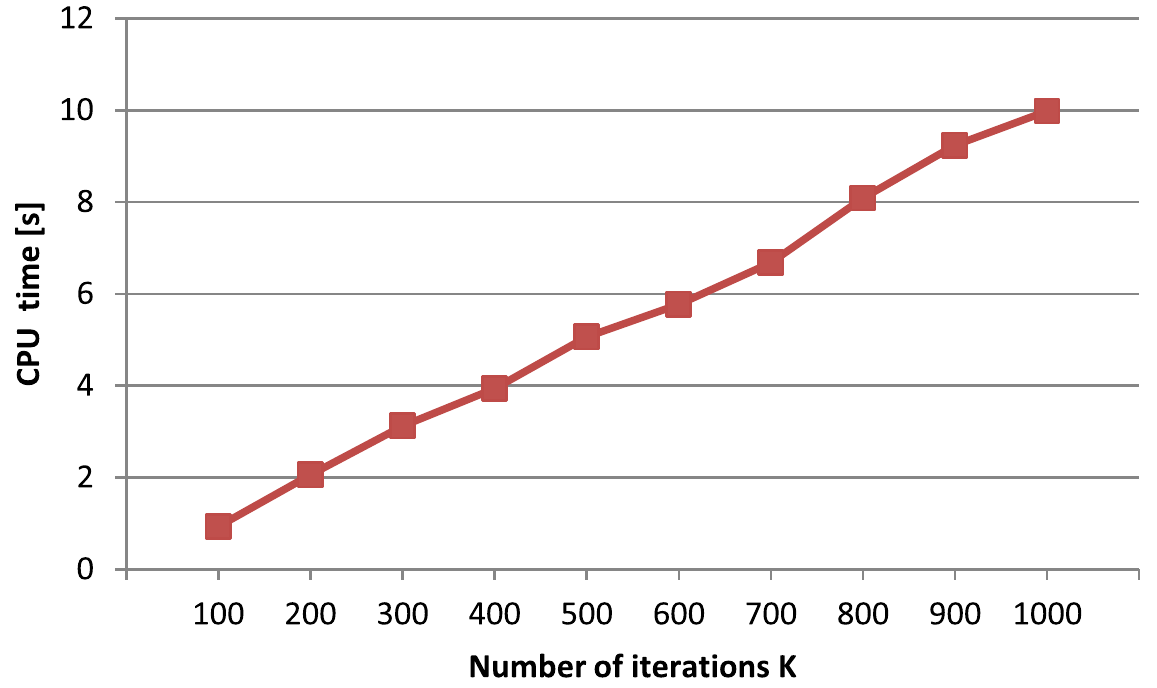}
}
\subfigure[\text{$K$ from 1000 to 10000.} ]{
\label{f:rrt.time.complexity.c2}
\includegraphics[width=0.45\textwidth]{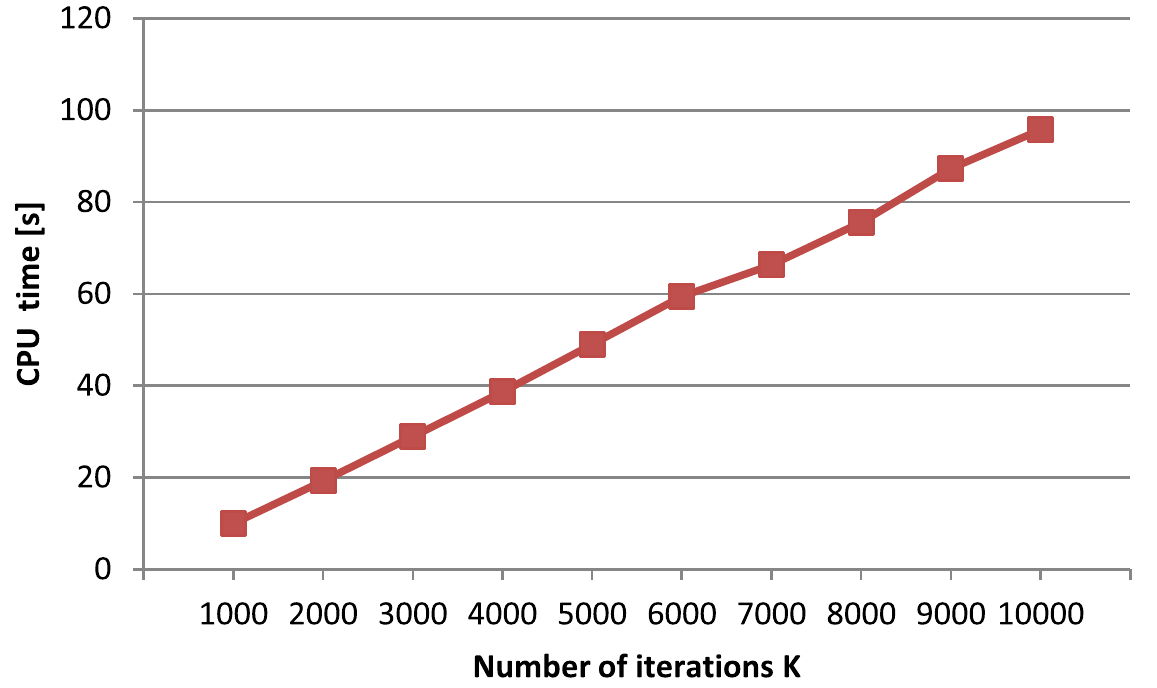}
}
\caption{Computation time for a continuous SMIB system.} \label{f:rrt.time.complexity.continuous}
\end{figure}

\begin{figure}[htb]
\centering
\subfigure[\text{$K$ from 100 to 1000.}]{
\label{f:rrt.time.complexity.h1}
\includegraphics[width=0.45\textwidth]{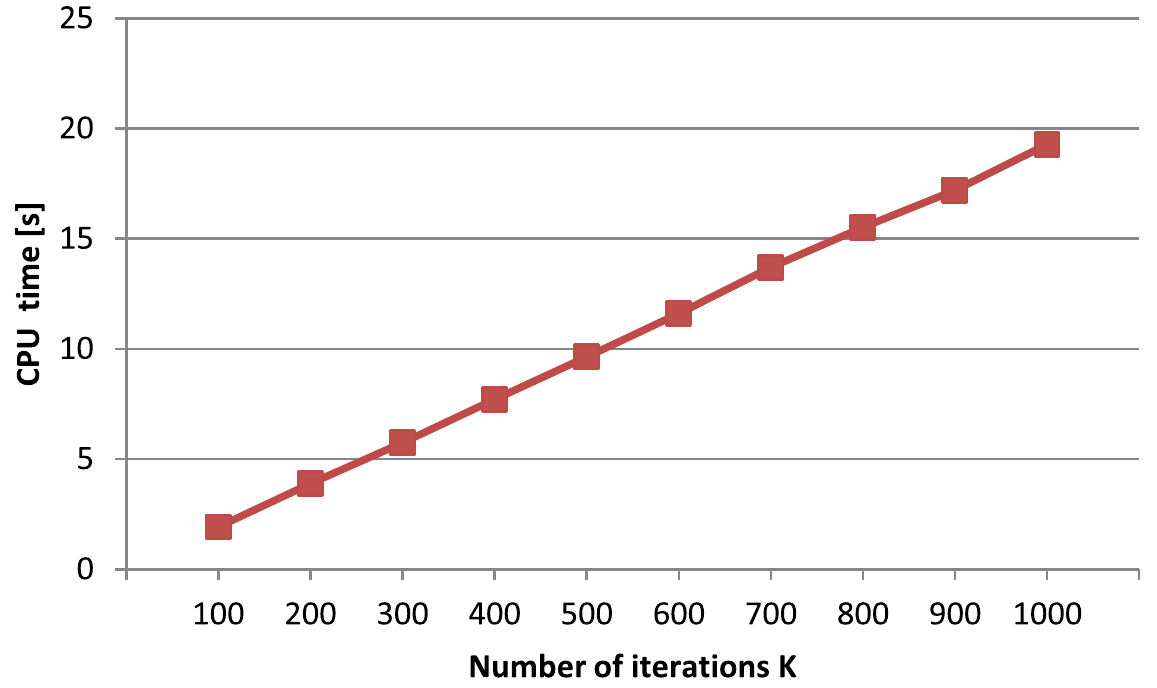}
}
\subfigure[\text{$K$ from 1000 to 10000.}]{
\label{f:rrt.time.complexity.h2}
\includegraphics[width=0.45\textwidth]{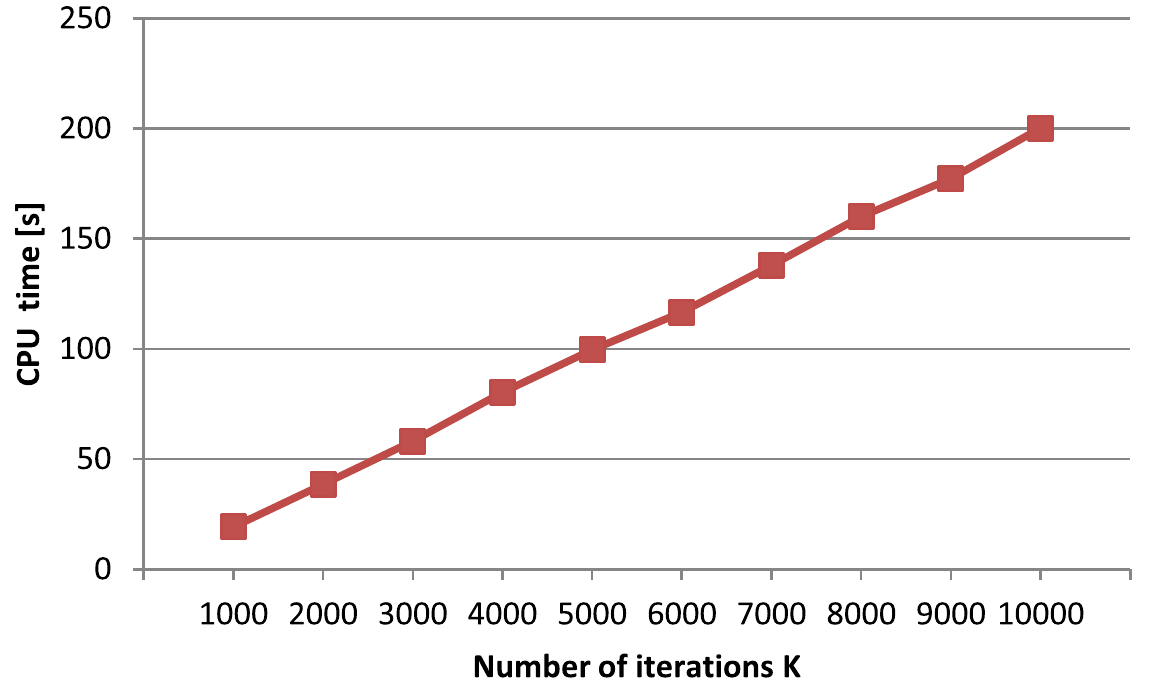}
}
\caption{Computation time for a hybrid SMIB system with two discrete states.} \label{f:rrt.time.complexity.hybrid}
\end{figure}

In practice, the simulation and collision detection in Step 4 cost much more time than other procedures, i.e., $c_e >> c_s + c_a + c_n$, resulting from the numerical integration and calculation conducted in this step.
So the actual computation time grows almost linearly with respect to the number of iterations $K$.
As for a demonstration, the computation time for constructing RRT searching trees for the continuous dynamics and hybrid dynamics with two discrete states of a SMIB system with different $K$ is shown in Fig. \ref{f:rrt.time.complexity.continuous} and Fig. \ref{f:rrt.time.complexity.hybrid}, respectively.
In all the continuous and hybrid cases, it is observed that the computation time for the RRT searching trees extension grows approximately linearly with $K$.
Settings in these computation are the same with Ref. \cite{wu2016}, and the computation is conducted with MALTAB 2010b on the Windows~7 32-bit OS
on a laptop with a intel$^{\textregistered}$ CORE$\texttrademark$ i7-2640M CPU @2.80GHz and 4G memory.

\section{Application to Power System Dynamic Security Analysis}
\label{sec:application}
In this section, an application to dynamic security analysis of the 39-bus New England power system \cite{athay1979,pai1989} exhibiting hybrid dynamics is employed to illustrate the effectiveness and performance of the proposed method, where the tool we adopt is PSAT for both modeling and simulation.

\begin{figure}[!t]
\center
\hspace{0.2in}
\includegraphics[width=0.9\textwidth]{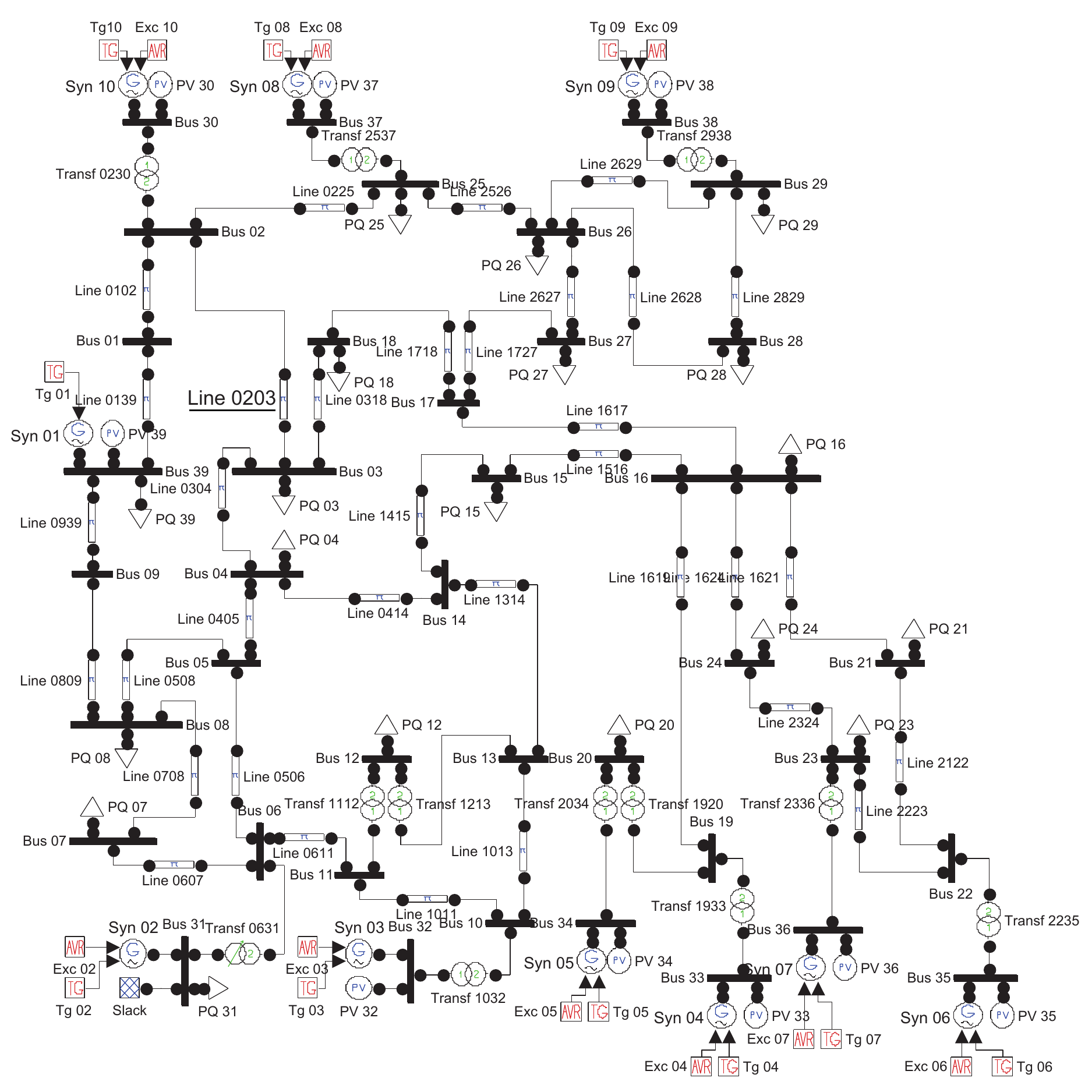}
\caption{The PSAT model of 39-bus New England power system
with two operation modes: Mode 1: Line 0203 open and Mode 2: Line 0203 closed.}\label{f:past.model.39bus}
\end{figure}

\subsection{Model of Hybrid Dynamics}
The 39-bus New England power system, also known as 10-machine system, is a widely-used benchmark representing the 345kV transmission system in the New England area of US, and its one-line diagram is shown in Fig. \ref{f:39bus}.
This system consists of 10 machines, 39 buses, 12 transformers and 34 lines, where Generator 1 (on Bus 39) represents the aggregation of a large number of generators outside New England.
The system data are taken from Ref. \cite{athay1979}.
Additional controllers including Turbine Governors (TGs) and Automatic Voltage Regulators (AVRs) are utilized in the model, with their parameters cited from Ref. \cite{bagheri2013}.

%
%
We consider a hybrid case for the 39-bus system including both clearing and re-closing operations on a faulted line (Line 0203: from Bus 2 to Bus 3).
The dynamics are represented with the following hybrid automaton $\mathbb{H}$:
\begin{itemize}
\item Discrete states: $Q = \{q_1, q_2\}$, where $q_1$ corresponds to the status Line 0203 open, and $q_2$ closed;

\item Continuous states: $X = \mathbb{R}^{106} \times \mathbb{R}^{137}$, and hereafter, the symbol $x$ denotes all the 106 state and 137 algebraic
variables, unless otherwise specified;

\item Control inputs: $U = \{   \pm 0.5,   \pm 0.4,   \pm 0.3,   \pm 0.2,   \pm 0.1,   0  \}$, and the control $u \in U$ is applied to the reference voltage, ${v_{\rm ref}}$, of AVR 8 on Generator 8;

\item Physical model: The PSAT model with a DAE implicitly built-in;

\item Initial state: A fault is defined on Bus 3 since $t=1$s, and is cleared at $t=1.1$s by disconnecting Line 0203 with a breaker, and the system data at $t=1.1$s is taken as the initial state $s_{\rm init}$, while the discrete state for $s_{\rm init}$ is $q_1$;

\item Edges: $E=\{(q_1,q_2), (q_2,q_1)\}$;

\item Guards: $G(q_1,q_2) = G (q_2,q_1) = X$; and

\item Reset maps: $R(q_1,q_2,x) =R(q_2,q_1,x) = x$.
\end{itemize}
This automaton has two admissible switching operations described by  the edges set  $E$ and the corresponding guards described by $G$.
The reset map in each switching is an identity, indicating that the switching of discrete state does not change the continuous states.

The PSAT model of the continuous dynamics in hybrid automaton $\mathbb{H}$ for this 39-bus system is shown in Fig. \ref{f:past.model.39bus},
where all the lines are connected.
In the figure, the component Syn $i$ is the $i$-th synchronous generator, Tg $i$ the turbine governor connected to the $i$-th generator, Exc $i$ the automatic voltage regulator connected to the $i$-th generator.
Notice that there are two controllers, one  Tg and one Exc, for every generator except the Syn 01, where there is only a Tg 01.
The components named Bus, Line, Transf, PV, PQ and Slack correspond to a bus, line, transformer, PV bus,  PQ bus and slack bus\footnote{
In power systems, a PV bus is also referred to as generator bus/node or voltage-controlled node, where the real power and the voltage amplitude are specified; a PQ bus is also called load bus/node, where the real power and the reactive power are known; a slack bus, also known as reference bus or swing bus, is used to balance the active and reactive power in the system \cite{machowski1997}. There is only one slack bus in one power system, and
both the voltage phase and amplitude are known for the slack bus.}, respectively.
The system model of continuous dynamics in $q_1$ is obtained by disconnecting the faulted line (Line 0203 shown with underline in Fig. \ref{f:past.model.39bus}).

\subsection{Dynamic Security Analysis with RRT Computation}
\begin{figure}[tbh]
\centering
\includegraphics[width=0.7\textwidth]{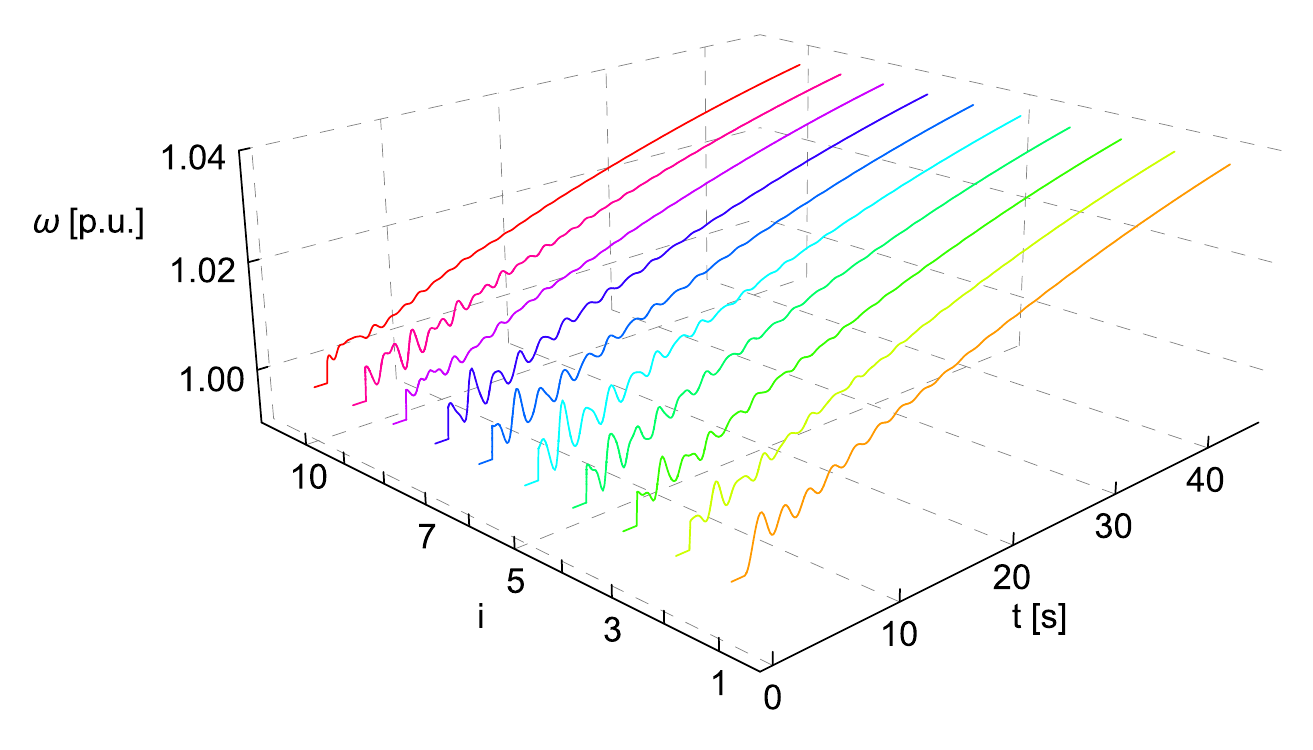}
\vspace{-0.1in}
\caption{Time responses of rotor speeds $\omega_i$, $i=1, \dots, 10$, in $q_1$ without proper control.}
\label{f:fault.response}
\end{figure}

Figure \ref{f:fault.response} shows the dynamic responses of rotor speeds of the pre-, on- and post-fault system for the benchmark.
In the pre-fault system, the generators are synchronous, as they work at the same rotor speed $\omega =1$.
The fault causes a significant increase in rotor speeds.
After the fault is cleared by disconnecting the faulted line, the rotor speeds oscillate for several seconds,
indicating the generators become asynchronous.
Though the difference of rotor speeds among generators decrease with time, the absolute rotor speeds keep increasing without proper control, i.e., $u=0$.
It may result in the tripping of all of the generators in the worst case scenario \cite{susuki2011}.

We hence perform dynamic security analysis for the 39-bus system to check whether it could attain its normal operation mode after the given fault.
In this case study, the target set $S_{\rm goal} $ is defined as a range of continuous variables in $q_2$:
\begin{eqnarray*}
S_{\rm goal}    =   \{(q_2, x) :&  |\omega_i - 1|  \leq  0.01,& i  =  1,  \dots,  10; \\
 &  |\theta_j - \theta_k |  \leq  \pi/6, & j,k =  1,  \dots,  39 ; \\
 &  |v_k - 1|  \leq  0.2, &  k  =  1,  \dots,  39 \},
\end{eqnarray*}
where $\omega_i$ in p.u. is the rotor speed of Generator $i$, $\theta_j$ in rad the voltage phase of Bus $j$, $v_k$ in p.u. the voltage amplitude of Bus $k$, and $| \cdot|$ the absolute value.
The setting of $\omega$ indicates the requirement on synchronization of generators, and the later two are to ensure stable power supply.

The switching between discrete states neither changes the continuous variables nor has time cost in our application.
Besides, the AVR reference voltages ${v_{\rm ref}}$ and the TG reference rotor speeds ${\omega_{\rm ref}}$ are constants during time-domain simulation.
It is thus reasonable to exclude these discrete and continuous variables in defining the distance function.
Among the continuous variables, the generator rotor angles $\delta$ and the bus-voltage phases $\theta$ belong to a circle, namely a nonlinear space, and each has a period of $2 \pi$.
Therefore, a hybrid distance for any two hybrid states $s=(q,x)$ and $s'=(q',x')$ is defined as follows:
\[
\rho(s,s') := \rho(x,x') \text{ and }
\rho(x,x'):=\rho(x_{\rm n},x'_{\rm n}) + \rho(x_{\rm l},x'_{\rm l}),
\]
where $x_{\rm n}$ represents the nonlinear variables $\delta$ and $\theta$, and $x_{\rm l}$ other continuous variables excluding $v_{\rm ref}$ and $\omega_{\rm ref}$.
For the linear part, the distance is
$$ \rho(x_{\rm l}, x'_{\rm l}) := \| x_{\rm l} - x'_{\rm l}\|_2,$$
where $\| \cdot\|_2$ is the $l_2$ norm of a vector, and the distance for the nonlinear variables is
\begin{eqnarray*}
\rho(x_{\rm n}, x'_{\rm n}) :=  \|( \rho(\delta_1, \delta_1'),\dots,\rho(\delta_{10}, \delta_{10}'),
\rho(\theta_1, \theta_1'),\dots,\rho(\theta_{39}, \theta_{39}') ~ )^{T} \|_2,
\end{eqnarray*}
where $^T$ stands for the transpose operation of vectors, and for any nonlinear variable, say $\delta_i$,
$$ \rho(\delta_i , \delta_i')  : = \min_{k \in \mathbb{Z}} | \delta_i - \delta_i' - 2 k \pi|. $$

The DAE models of continuous dynamics characterize the manifolds for system trajectories.
Searching in the whole hybrid state space, $S = Q \times X$, seems inefficient and impractical.
But, it is also challenging to search on the manifolds as in Ref. \cite{xu2010}, due to lack of structure information and high dimension of the problem.
As an alternative, we propose to search in some near-by area containing (parts of) the manifolds, and adopt the uniform sampling strategy in the search space.
The search space is $Q$ for the discrete variable, and it is selected for the continuous variables excluding $v_{\rm ref}$ and $\omega_{\rm ref}$ following the three principles:
\begin{itemize}
\item The ranges are $(- \pi, \pi]$ for the nonlinear variables.

\item The ranges are $[x_{\rm min},x_{\rm max}]$ for the linear variables with predefined boundaries in system model.

\item  The ranges are $[~x - \Delta x , x + \Delta x ~]$, otherwise, where $ \Delta x  = 0.1  |x^*|$, and $x^*$ is the value of $x$ in the equilibrium point obtained in the power flow computation with basic model. (With abuse of notation, the symbol $x$ here represents a scalar variable.)
\end{itemize}
\begin{figure}[!htb]
\centering
\subfigure[\text{[0, 300] iterations.}]{
\label{f:rrt.tree.extension.sub1}
\includegraphics[width=0.6\textwidth]{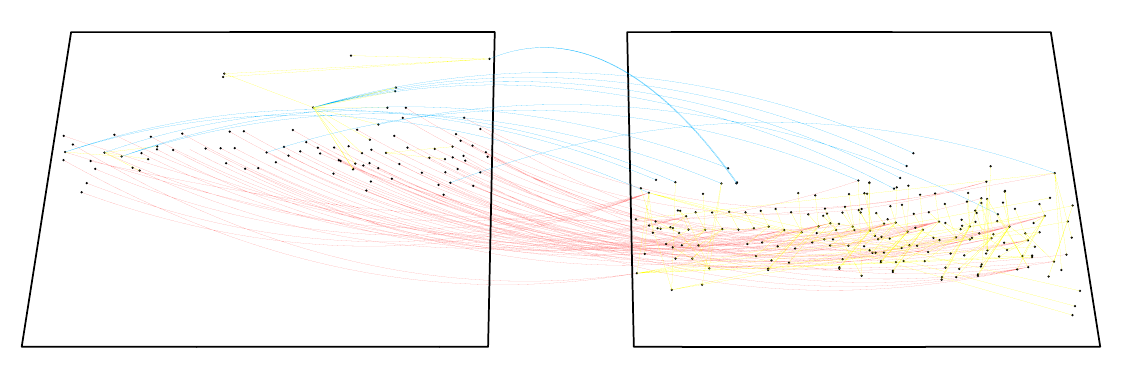}
}
\subfigure[\text{[301, 600] iterations.}]{
\label{f:rrt.tree.extension.sub2}
\includegraphics[width=0.6\textwidth]{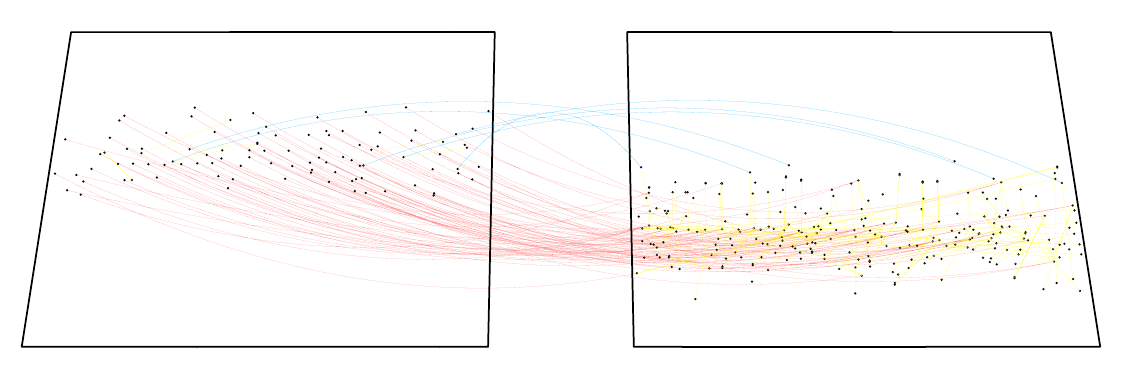}
}
\subfigure[\text{[601, 900] iterations.}]{
\label{f:rrt.tree.extension.sub3}
\includegraphics[width=0.6\textwidth]{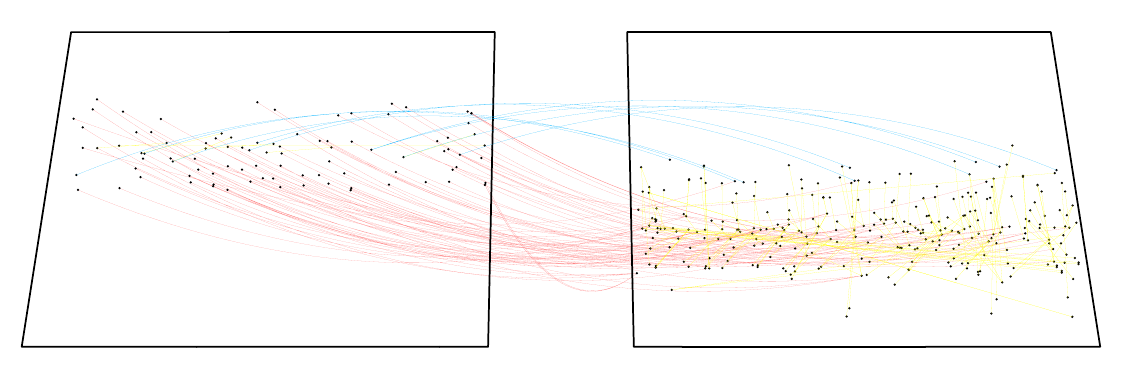}
}
\subfigure[\text{[901, 1200] iterations.}]{
\label{f:rrt.tree.extension.sub4}
\includegraphics[width=0.6\textwidth]{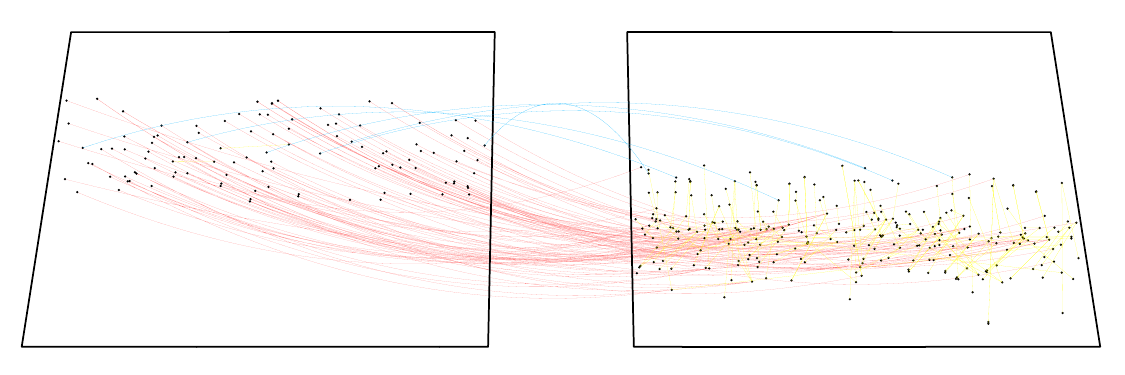}
}
\subfigure[\text{[1201, 1500] iterations.}]{
\label{f:rrt.tree.extension.sub5}
\includegraphics[width=0.6\textwidth]{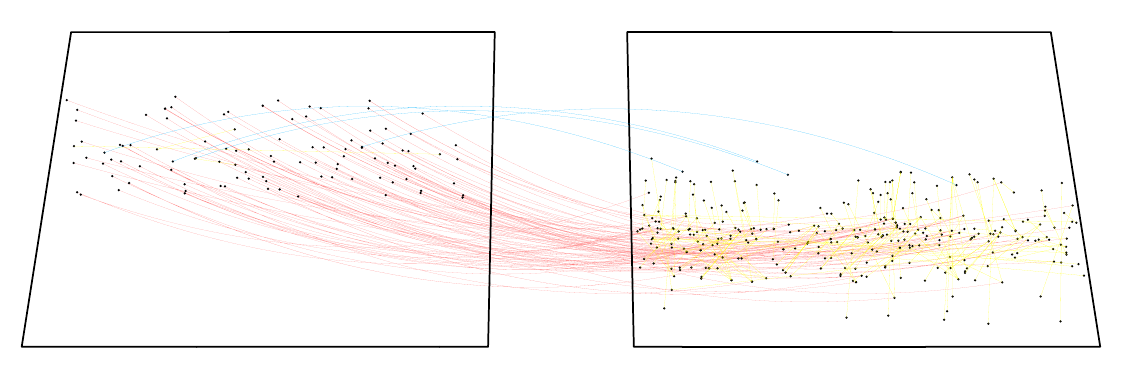}
}
\caption{Extension of RRT in hybrid state space.
The left panel corresponds to the discrete state $q_1$, and the right $q_2$.
The black dots are the nodes of the searching tree.
The light blue edges are from $q_1$ to $q_2$, and the light red edges are from $q_2$ to $q_1$.
The yellow edges specify the continuous evolution in the same discrete state.
The density of nodes grows with number of iterations.
} \label{f:rrt.tree.extension.p1}
\end{figure}

\begin{figure}[!t]
\centering
\includegraphics[width=0.67\textwidth]{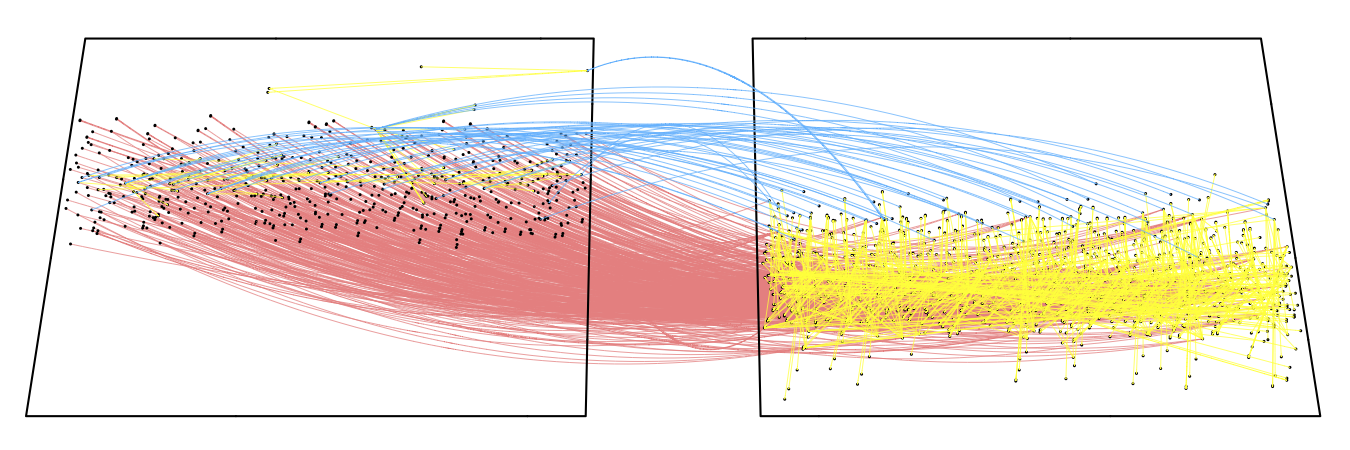}
\text{$q_1$  \quad\quad\quad\quad\quad\quad\quad\quad\quad\quad\quad\quad\quad\quad\quad  $q_2$}
\caption{The constructed RRT searching tree in hybrid state space with $K=2000$ and $\Delta t = 1.26$s.} \label{f:rrt.tree.extension.p2}
\end{figure}

\begin{figure}[!b]
\centering
\subfigure[Rotor speeds $\omega_i$, $i=1, \dots, 10$.]{
\label{f:result.sub.o}
\hspace{0.1in}
\includegraphics[width=0.7\textwidth]{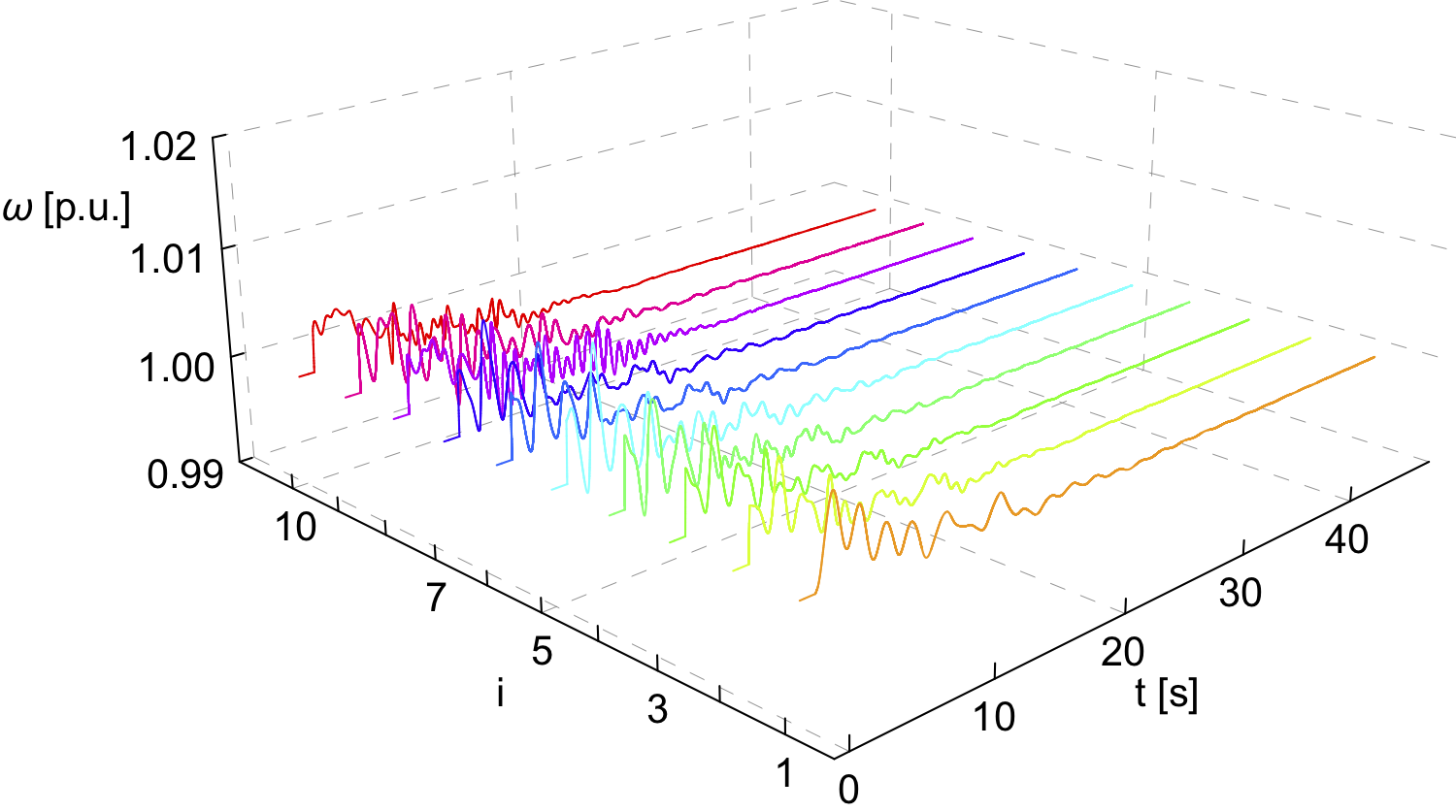}
}\\
\subfigure[Relative rotor angles with respect to the Center-Of-Inertia, $\delta_i-\delta_{\rm COI}$, $i=1, \dots, 10$.]{
\label{f:result.sub.d}
\hspace{-0.11in}
\includegraphics[width=0.7\textwidth]{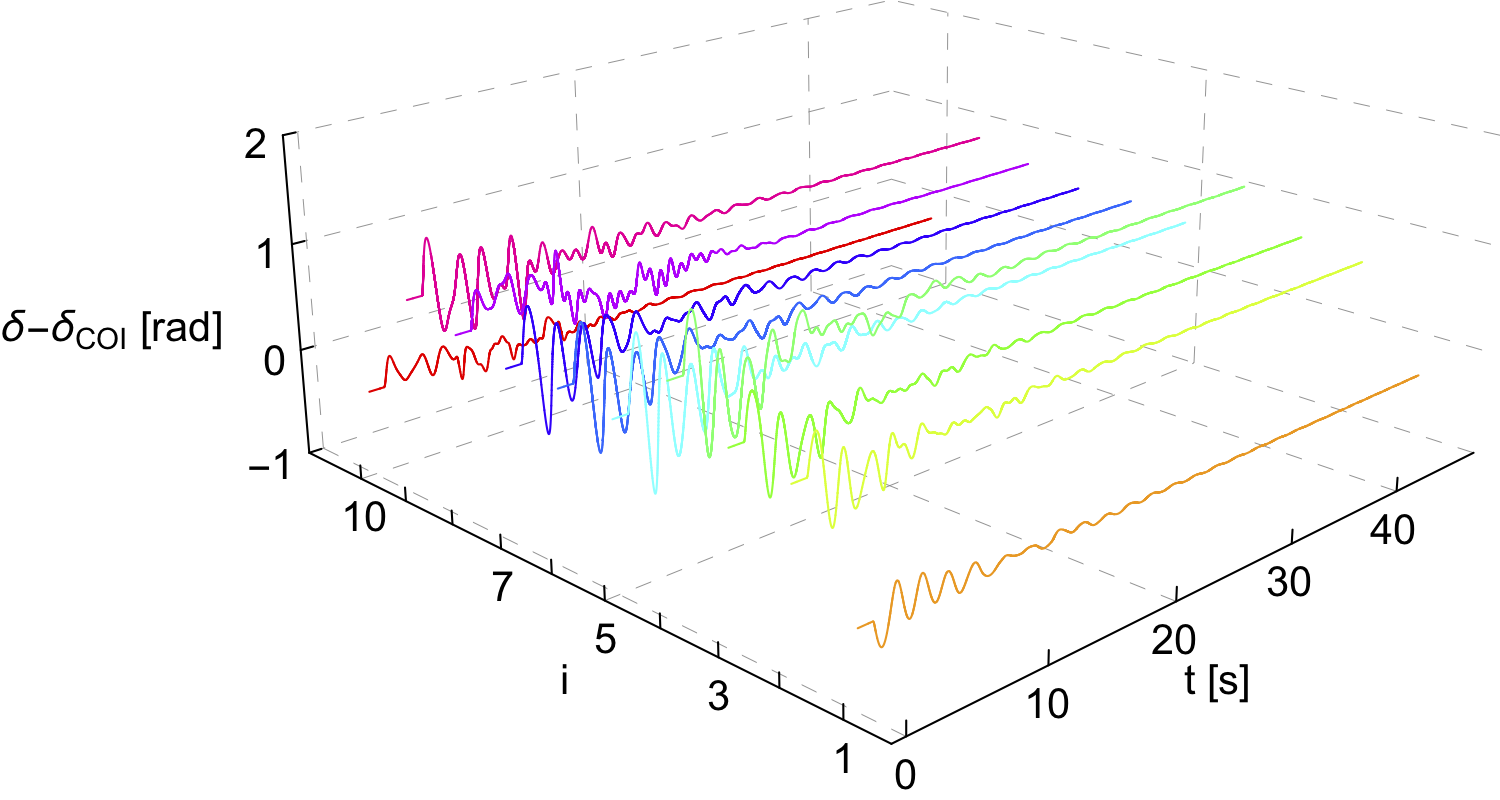}
}
\caption{Time responses of variables $\omega$ and $\delta$ for generators along the execution.}\label{f:result.g}
\end{figure}

The RRT computation is carried out in this hybrid case of 39-bus system using the proposed framework in Sec. \ref{sec:sampling.based.algorithms}.
Figure \ref{f:rrt.tree.extension.p1} shows the extension of a RRT searching tree $\mathcal{T}$ in the hybrid state space.
The projection of the finally constructed searching tree $\mathcal{T}$ on $\delta_8-\omega_8$ planes with two discrete states is shown in Fig. \ref{f:rrt.tree.extension.p2}, where $\mathcal{T}$ is constructed by using the sampling time interval $\Delta t =1.26$s and number of iterations $K=2000$.
The setting of $\Delta t$ is the stabilizer time constant of AVR 8, while the choice of $\delta_8-\omega_8$ planes is related to Generator 8
with control input.

\begin{figure}[!t]
\centering
\subfigure[Relative bus-voltage phases with respect to their spatial center, $\theta_j - \theta_{\rm avg}$, $j =1, \dots, 39$.]{
\label{f:result.sub.t}
\hspace{-0.2in}
\includegraphics[width=0.7\textwidth]{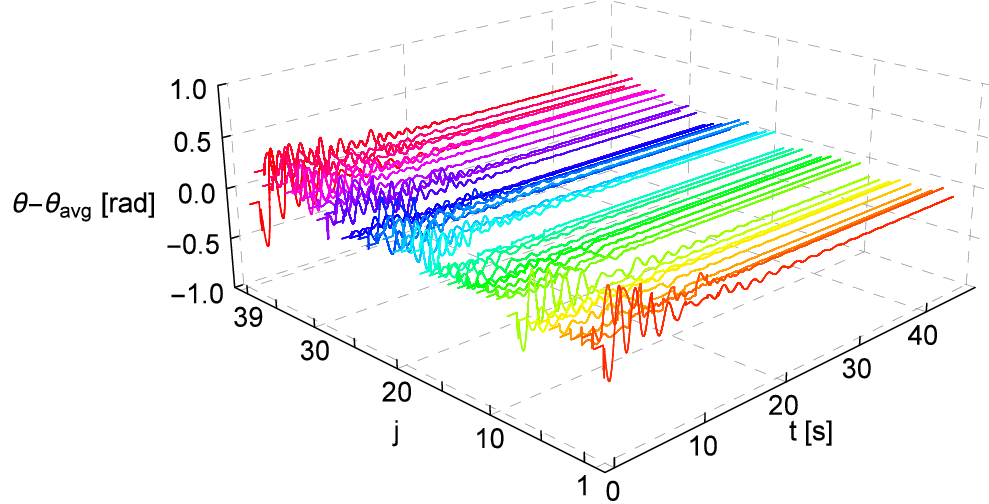}
}\\
\subfigure[Bus-voltage amplitudes $v_j$, $j=1, \dots, 39$.]{
\label{f:result.sub.v}
\hspace{0.2in}
\includegraphics[width=0.7\textwidth ]{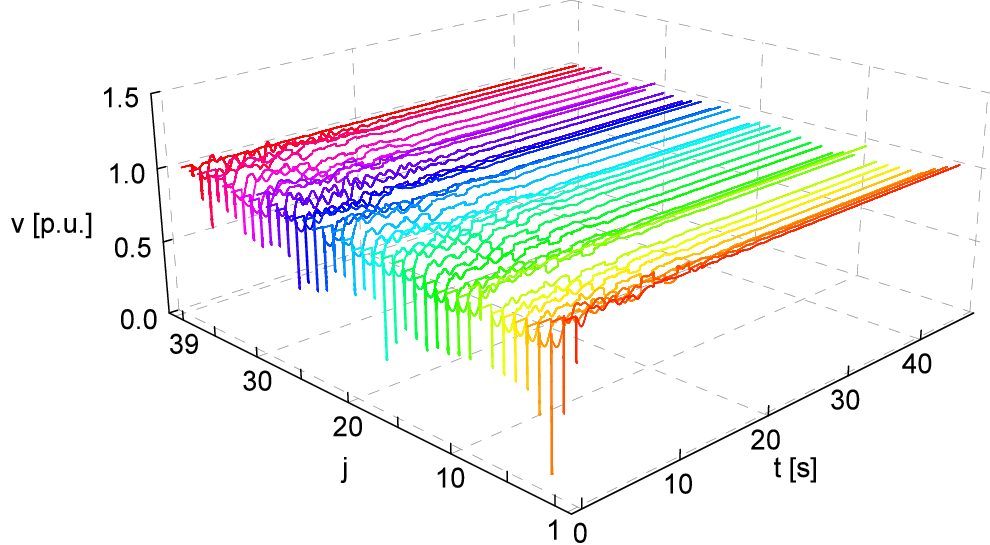}
}
\caption{Time responses of variables $\theta$ and $v$ for buses along the execution.} \label{f:result.b}
\end{figure}

\begin{figure}[tb]
\centering
\includegraphics[width=0.7\textwidth]{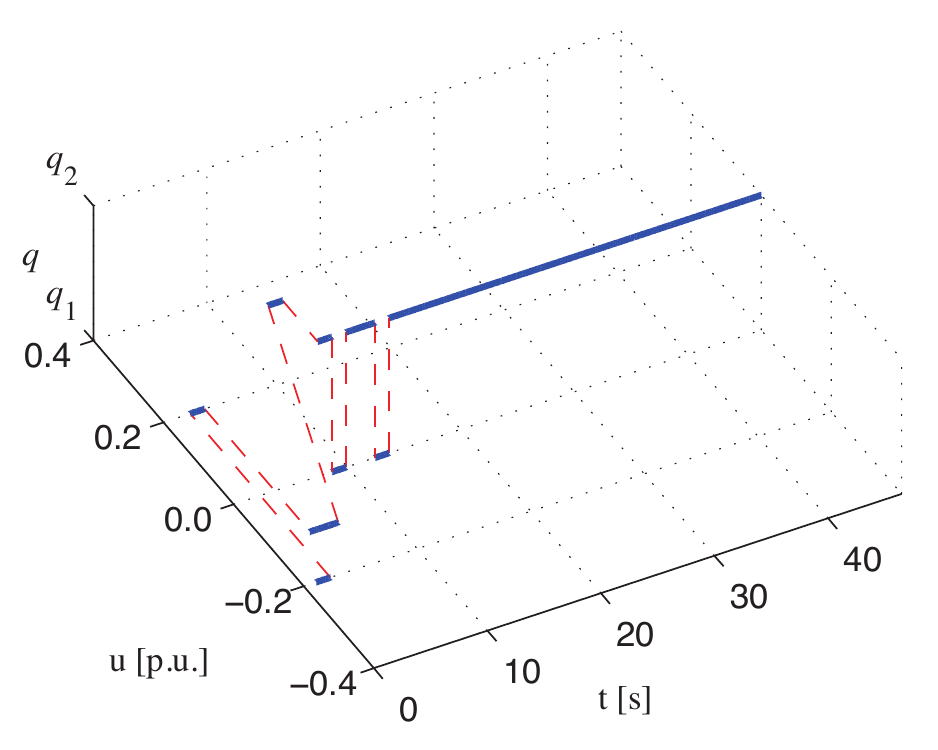}
\caption{The discrete state and input sequence along the feasible execution.}\label{f:result.qu.3d}
\end{figure}

\begin{figure}[!h]
\centering
\hspace{-0.7in}
\subfigure[Discrete state $q$.]{
\label{f:result.sub.q}
\includegraphics[width=0.45\textwidth]{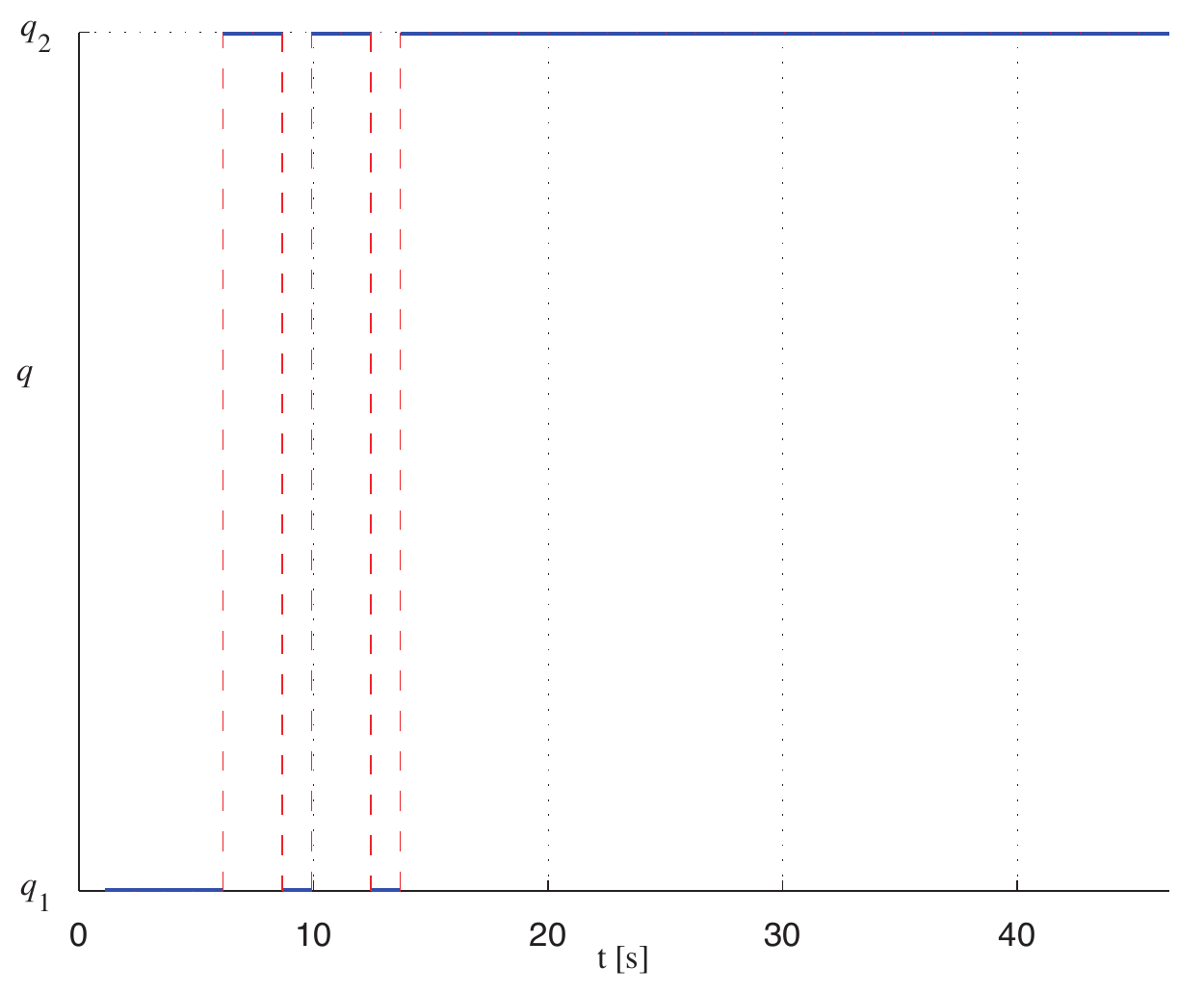}
}
\hspace{-0.3in}
\subfigure[Input sequence $u$.]{
\label{f:result.sub.u}
\includegraphics[width=0.45\textwidth]{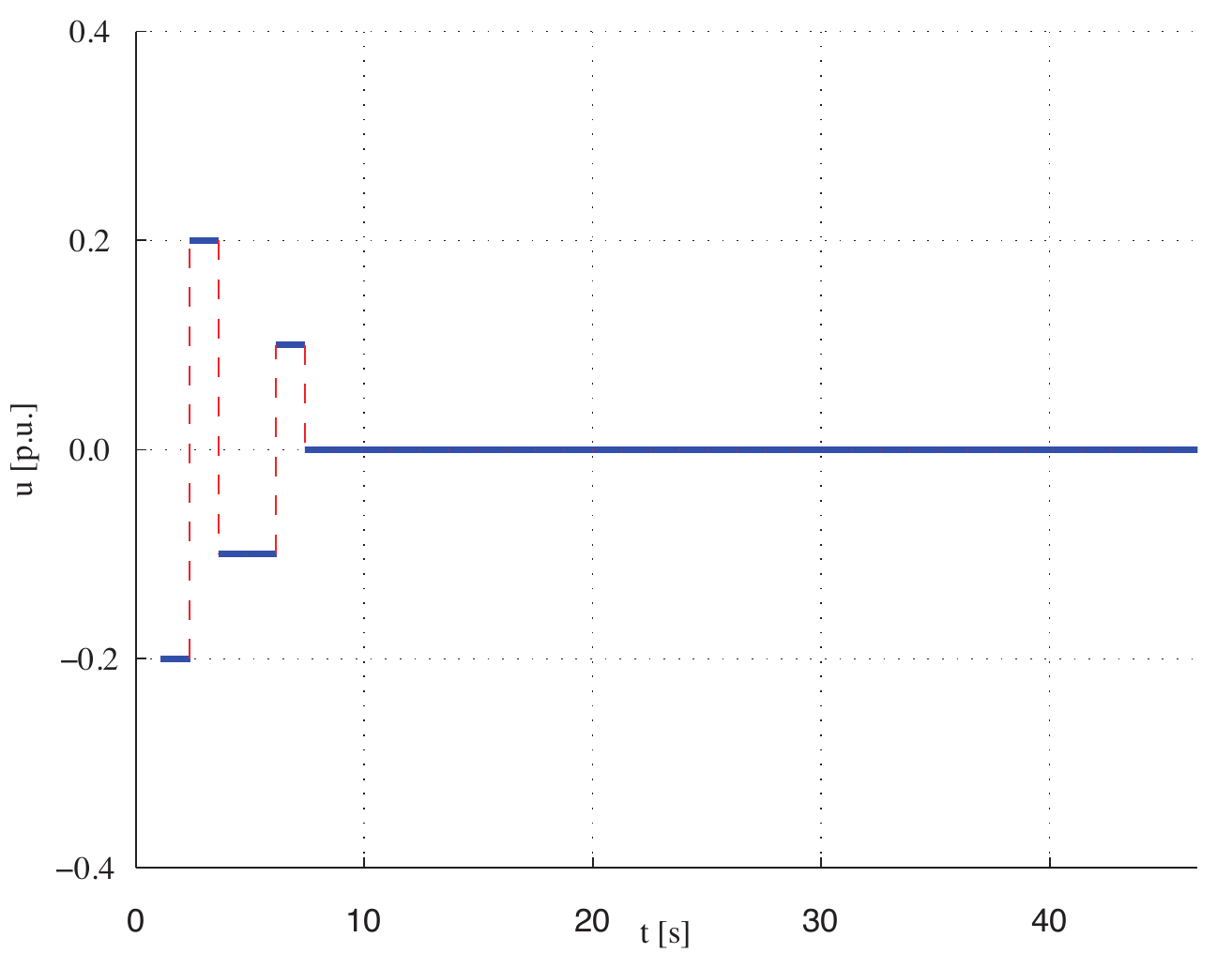}
}
\caption{Time responses of discrete state $q$, and the input sequence along the feasible execution.} \label{f:result.qu}
\end{figure}

A feasible execution is obtained on the tree $\mathcal{T}$ shown in Fig. \ref{f:rrt.tree.extension.p2} and the depth of this execution is $36$.
Figures \ref{f:result.g}-\ref{f:result.b} show time responses of continuous variables $\omega$, $\delta$, $\theta$ and $v$ along the execution.
As we have observed in Fig. \ref{f:fault.response}, the rotor speeds $\omega$ of generators soon increase after the fault and the whole system will lose its stability without proper control actions.
But, on the obtained execution shown in Fig. \ref{f:result.sub.o}, even though they initially oscillate significantly after the fault,
the rotor speeds finally converge to one value (near $1$) with the increase of time, indicating the generators reach a (new) synchronous state.
The dynamic responses of rotor angles $\delta$ are shown in Fig. \ref{f:result.sub.d}.
As an alternative to the original (or absolute) $\delta $, the relative rotor angles with respect to the \textit{Center-Of-Inertia} \cite{kundur1994}, $\delta  - \delta_{\rm COI}$, are shown in the figure for the generators.
The center of inertia is defined by
\begin{eqnarray*}
\delta_{\rm COI} := \sum_{i=1}^{10} \frac{H_i}{H} \delta_i,
\end{eqnarray*}
where $H_i$ is an inertia constant and $H = \sum_{i=1}^{10}  H_i $.
The data for generator inertia constants are taken from Ref. \cite{pai1989}.
It is shown in Fig. \ref{f:result.sub.d} that all the relative rotor angles eventually converge along the execution.
This observation coincides with the convergence of $\omega$ in Fig. \ref{f:result.sub.o}.
The relative rotor angles tend to different values for different generators, as the generators initially have different phases before the fault.
Figure \ref{f:result.b} shows that the system could provide stable power supply along the searched execution with proper control actions.
Time responses of bus-voltage phases $\theta$ are shown in Fig. \ref{f:result.sub.t}, where we display the relative phases with respect to their spatial average
\begin{eqnarray*}
\theta_{\rm avg} := \frac{1}{39}\sum_{j=1}^{39}  \theta_j,
\end{eqnarray*}
in accordance with the requirement on $\theta$ for the target set $S_{\rm goal}$.
Figure \ref{f:result.sub.v} is for the bus-voltage amplitudes, and it indicates that the bus-voltages are affected significantly by the fault,
but recover their stability along the execution.
Figure \ref{f:result.qu.3d} is the input sequence and discrete switching along the execution.
Figure \ref{f:result.sub.q} shows the discrete switching of the system between the two states.
Figures \ref{f:result.g}-\ref{f:result.b} and Fig. \ref{f:result.sub.q} show that both requirements for the dynamic security analysis defined in Problem \ref{p:DSA} are fulfilled, and thus verify the dynamic security of the system with respect to the given fault.

In addition to verifying the dynamics security, the feasible execution also synthesizes a concrete control strategy to drive the system to run in a target area while maintaining the security.
Figure \ref{f:result.sub.q} indicates the discrete controls to switch between the operation modes, while Fig. \ref{f:result.sub.u} shows the continuous control input sequence in each operation mode along the execution.
Note that in Fig. \ref{f:result.sub.q} multiple times of line switching are observed.
Normally, such multiple switching are avoided in practice because they would cause stressful damage of the transmission systems.
But, as shown in Ref. \cite{susuki2012}, utilizing the idea of line switching effectively provides a new possibility of power systems control to keep safety operation.
The demonstrated operation in Fig.  \ref{f:result.sub.q} is another example of the effective line switching protocol for the safe operation.
Compared with other control strategies, this obtained strategy provides a guaranteed solution and a practical approach to large-scale problems.

\subsection{Discussion on Computation Efficiency and Application}
In conducting the demonstration, the computation environment is MALTAB 2010b working on the Windows~7 32-bit system, and the hardware platform is a laptop with an intel$^{\textregistered}$ CORE$\texttrademark$ i7-2640M CPU @2.80GHz and 4G memory.
As discussed earlier, the modeling tool and simulator for power system application is the PSAT toolbox version 2.1.6.
The computation time is $27345$s to build this searching tree with $\Delta t =1.26$s and $K=2000$.
The computation is conducted in a serial manner for this problem with $243$ continuous variables,
$2$ discrete states and $11$ control inputs, which means, in each loop of the RRT computation,
$2 \times 11$ times simulations are carried out one-by-one for a DAE with $243$ equations.
The computation time to conduct a single dynamic simulation is around $1.6896$s for this 39-bus system.
Therefore, the total computation time would be no less than $74432.4$s by using the classical RRT framework.
The adopted method of reusing simulation results saves at least $63\%$ of the computation time, since several nodes may share the same parent node in practice.

The current computation time in the demonstration is relatively long, compared with the execution time.
This is due to the time domain simulation and incremental search in the method.
The computation time of the proposed method can be reduced with different approaches.
Since the simulation procedure is recognized as the bottleneck of the computation in Sec. \ref{sec:time.complexity} and the time-domain simulations from one nearest node with different pairs $(q_i,u_j)$ are totally independent, the computation time can be reduced significantly by adopting parallel computing.
In addition to parallelization, Field-Programmable Gate Array (FPGA) based hardware acceleration of the dynamic simulation of power systems \cite{yang2013} could also be used to improve the computation efficiency.

Regarding the application of the proposed method, the post-contingency on-line use is not suitable at the current stage.
From the power grid perspective, proper emergency control actions are required to be derived and taken within several seconds to prevent the spreading of the fault over the system networks and to stabilize the grid, just after a contingency really happens.
Considering its computation time, the proposed method has not been suitable at the current stage for on-line applications of large-scale systems when a contingency really occurs.

Even though, it should be emphasized that the proposed security analysis method has a clear meaning and practical values in the off-line use.
For example, the following two use cases can be considered for the off-line application scenarios of the proposed method.

In standard dynamic security assessment systems \cite{chiang2010}, it is a common idea to obtain a ranked list of potentially harmful contingencies periodically from a set of possible contingencies based on the real-time data of power systems by using the dynamic contingency screening and selection module.
Time-domain simulation is further conducted on the selected potential contingencies to evaluate the stability of the system.
Therefore, one use case of the off-line application scenario is to generate a proper set of the system's behavior with control actions against the set of contingencies, in order to derive a preventive control action.
The proposed approach is suitable to this off-line use case with strong advantage that it is capable of deriving a provably-correct action via reachability analysis, which has not been conducted in power system control research.
Limitations of this off-line use lie in the power of representative model for the system's behavior and the choice of set of contingencies.
For example, if a target practical grid exhibits transient dynamics due to a contingency that is not included in the set, then the proposed approach does not generate a provably-correct control action.

Another use case of the off-line application scenario is to generate a dynamically-relevant dataset for training a model in the machine learning-based method \cite{shenoy2016} for DSA.
Compared to the dataset from the physical world, the dataset from simulation is easier to access, in addition to the flexibility of considering more operation modes of the system.
Limitations of this off-line use lie in the representativeness of the dataset and the choice of the machine learning model.
Even though dataset obtained using the proposed approach is provably-correct, the correctness of the trained model may not be guaranteed, due to the probabilistic nature of the machine learning methods.

\section{Conclusion}
\label{sec:conclusion}
Security is one of the main concerns in power systems on providing continuous and stable power supply service.
Since power systems are typical CPS, their dynamics possess strong hybrid nature resulted from the interaction of the cyber- and physical worlds.
The hybrid nature and the high-dimension properties of the dynamics make the analysis to the security of power systems a challenging problem.

This paper proposed to deal with the dynamic security analysis of power systems by checking the reachability of the associated hybrid models, and focused on the index-1 DAE models regarding the continuous dynamics.
A sampling-based algorithm was developed to conduct the analysis in high-dimensional hybrid state space by extending the RRT algorithm on its sampling strategy, distance function, and computation framework.
With the adoption of a power system analysis toolbox called PSAT, the developed computation framework was applied to the dynamic security analysis problem of power systems.
A case study with a hybrid case on the 39-bus New England System was carried out to demonstrate the effectiveness and performance of the proposed approach.
In addition to the advantages of the RRT based computation as discussed in Ref. \cite{wu2016}, this architecture provides a practical approach to large-scale problems, especially for those with many continuous variables.
Since the adopted tool can be replaced with any other interested one in specified application domains, this method is feasible to handle problems in different fields with a wide variety of models.

In the past few years, cyber-attacks against power systems, especially the cyber network, have been manipulating the measurements and injecting false information to fail the critical modules, such as state estimation and Optimal Power Flow (OPF) in the power system.
For the cyber-security issue, the proposed method in this paper is capable of including an additional module that monitors and records the estimated state data, and detects a manipulation (“anomaly”) by comparing a difference in data.
This comes directly from our original idea because we have developed a module-based DSA framework that enables the use of any existing power system simulators, which was PSAT in the paper.
This would not be achieved in any existing DSA methods.
Hence, we speculate that the proposed method is firstly able to handle cyber-attacks in the modern DSA technology.
To address the issue about cyber-attacks is an important topic that follows the work presented in the paper.

Regarding the follow-up studies, we also plan to integrate the time-domain simulation based method of this work with practical direct methods, for example, the BCU method, and machine learning to develop a framework for the dynamic security assessment.


%

\section*{Appendix}
Some notations used in the paper are concluded in the following table.

\begin{center}
\centering
\makebox[0pt][c]{\parbox{1.0\textwidth}{%
    \begin{minipage}[b]{0.45\hsize}\centering
       \begin{supertabular}{c|l}
        \toprule
          Notations  & Descriptions \\
        \cmidrule{1-2}
          $\mathbb{R}^n$  & n-dimensional real space \\
          $\mathbb{N}_+$ & natural number space \\
          $\mathbb{H}$ & hybrid automaton \\
          $f$ & vector field \\
          $m, n , k$ & integers \\
          $q$ & discrete state \\
          $s$ & hybrid state \\
          $t, \tau, T$ & time \\
          $u$ & control input \\
          $v$  & vertex; bus voltage amplitude \\
          $x$ & continuous state  \\
          $y$ & differential variable \\
          $z$ & algebraic variable \\
          $\rho$ & distance \\
          $\delta$ & generator rotor angle \\
          $\omega$ & generator rotor speed \\
          $\theta$ & bus voltage phase \\
        \bottomrule
        \end{supertabular}
    \end{minipage}
    \hfill
    \begin{minipage}[b]{0.45\hsize}\centering
       \begin{supertabular}{c|l}
        \toprule
          Notations  & Descriptions \\
        \cmidrule{1-2}
          $E$ & set of edges \\
          $F, \phi, \psi$ & vector-valued functions \\
          $G$ & guard condition \\
          $H$ & inertia constant \\
          $I$ & time interval \\
          $K$ & number of iterations \\
          $R$ & reset map \\
          $Q$ & set of discrete states \\
          $S$ & set of hybrid states \\
          $U$ & set of control inputs \\
          $X$ & set of continuous states \\
          $2^X$ & power set of $X$ \\
          $\mathrm{det}$ & determinate \\
          $\chi$ & execution \\
          $\mathcal{T}$ & searching tree \\
          $\mathsf{P}$ & probability \\
          $Reach$ & reachable set \\
        \bottomrule
        \end{supertabular}
    \end{minipage}%
}}
\end{center}

\bibliographystyle{IEEEtran}
\bibliography{./bib/implicit_hybrid_modeling_update_20180514}

\begin{thebibliography}{10}
\providecommand{\url}[1]{#1}
\csname url@samestyle\endcsname
\providecommand{\newblock}{\relax}
\providecommand{\bibinfo}[2]{#2}
\providecommand{\BIBentrySTDinterwordspacing}{\spaceskip=0pt\relax}
\providecommand{\BIBentryALTinterwordstretchfactor}{4}
\providecommand{\BIBentryALTinterwordspacing}{\spaceskip=\fontdimen2\font plus
\BIBentryALTinterwordstretchfactor\fontdimen3\font minus
  \fontdimen4\font\relax}
\providecommand{\BIBforeignlanguage}[2]{{%
\expandafter\ifx\csname l@#1\endcsname\relax
\typeout{** WARNING: IEEEtran.bst: No hyphenation pattern has been}%
\typeout{** loaded for the language `#1'. Using the pattern for}%
\typeout{** the default language instead.}%
\else
\language=\csname l@#1\endcsname
\fi
#2}}
\providecommand{\BIBdecl}{\relax}
\BIBdecl

\bibitem{cpsgroup2008}
{CPS Steering Group}, \emph{Cyber-Physical Systems Executive Summary},
  Arlington, Mar. 2008.

\bibitem{poovendran2010}
R.~Poovendran, ``Cyber-physical systems: Close encounters between two parallel
  worlds,'' \emph{P. IEEE}, vol.~98, no.~8, pp. 1363--1366, Aug. 2010.

\bibitem{nist2009}
{Electric Power Research Institute}, \emph{Report to NIST on the Smart Grid
  Interoperability Standards Roadmap (Contract No. SB 1341-090CN-0031)}, Jun.
  17 2009.

\bibitem{esfahani2010}
P.~M. Esfahani, M.~Vrakopoulou, K.~Margellos, J.~Lygeros, and G.~Andersson,
  ``Cyber attack in a two-area power system: Impact identification using
  reachability,'' in \emph{Proc. American Control Conference}, Jun. 2010, pp.
  962--967.

\bibitem{athay1979}
T.~Athay, R.~Podmore, and S.~Virmani, ``A practical method for the direct
  analysis of transient stability,'' \emph{IEEE Trans. Power App. Syst.}, vol.
  PAS-98, no.~2, pp. 573--584, Mar./Apr. 1979.

\bibitem{pai1989}
M.~A. Pai, \emph{Energy Function Analysis for Power System Stability}.\hskip
  1em plus 0.5em minus 0.4em\relax Dordrecht: Kluwer Academic, 1989.

\bibitem{chiang1995}
H.-D. Chiang, C.-C. Chu, and G.~Cauley, ``Direct stability analysis of electric
  power systems using energy functions: {Theory}, applications, and
  perspective,'' \emph{P. IEEE}, vol.~83, no.~11, pp. 1497--1529, Nov. 1995.

\bibitem{yang2014}
Q.~Yang, J.~Yang, D.~An, N.~Zhang, and W.~Zhao, ``On false data-injection
  attacks against power system state estimation: {Modeling} and
  countermeasures,'' \emph{IEEE Trans. Parallel Distrib. Syst.}, vol.~25,
  no.~3, pp. 717--729, Mar. 2014.

\bibitem{liang2017}
G.~Liang, J.~Zhao, F.~Luo, S.~R. Weller, and Z.~Y. Dong, ``A review on false
  data injection attacks against modern power systems,'' \emph{IEEE Trans.
  Smart Grid}, vol.~8, no.~4, pp. 1630--1638, Jul. 2017.

\bibitem{yang2017a}
Q.~Yang, D.~An, R.~Min, W.~Yu, X.~Yang, and W.~Zhao, ``On optimal pmu
  plancement-based defense against data integrity attacks in smart grid,''
  \emph{IEEE Trans. Inf. Forensics Security}, vol.~12, no.~7, pp. 1735--1750,
  Jul. 2017.

\bibitem{yang2017b}
Q.~Yang, D.~Li, W.~Yu, Y.~Liu, D.~An, X.~Yang, and J.~Lin, ``Toward data
  integrity attacks against optimal power flow in smart grid,'' \emph{IEEE
  Internet Things J.}, vol.~4, no.~3, pp. 1726--1738, Oct. 2017.

\bibitem{li2017}
Z.~Li, M.~Shahidehpour, and F.~Aminifar, ``Cybersecurity in distributed power
  systems,'' \emph{P. IEEE}, vol. 105, no.~7, pp. 1367--1388, Jul. 2017.

\bibitem{taskforce2003}
{U.S.-Canada Power System Outage Task Force}, \emph{Interim Report: Causes of
  the August 14th Blackout in the United States and Canada}, Nov. 2003.

\bibitem{poulsen2004}
\BIBentryALTinterwordspacing
K.~Poulsen. (2004) Software bug contributed to blackout. [Online]. Available:
  \url{https://www.securityfocus.com/news/8016}
\BIBentrySTDinterwordspacing

\bibitem{nasa2007}
K.~Morison, L.~Wang, and P.~Kundur, ``Powerless: Northeast blackout of 2003,''
  \emph{System Failure Case Study}, vol.~1, no.~10, pp. 1--4, Dec. 2007.

\bibitem{kim2016}
\BIBentryALTinterwordspacing
K.~Zetter. (2016) Inside the cunning, unprecedented hack of ukraine's power
  grid. [Online]. Available:
  \url{https://www.wired.com/2016/03/inside-cunning-unprecedented-hack-ukraines-power-grid/}
\BIBentrySTDinterwordspacing

\bibitem{andy2017}
\BIBentryALTinterwordspacing
A.~Greenberg. (2017) How an entire nation became russia's test lab for
  cyberwar. [Online]. Available:
  \url{https://www.wired.com/story/russian-hackers-attack-ukraine/}
\BIBentrySTDinterwordspacing

\bibitem{fink1978}
L.~H. Fink and K.~Carlsen, ``Operating under stress and strain,'' \emph{IEEE
  Spectrum}, vol.~15, no.~3, pp. 48--53, Mar. 1978.

\bibitem{wu1983}
F.~F. Wu and Y.-K. Tsai, ``Probabilistic dynamic security assessment of power
  systems-{I}: Basic model,'' \emph{IEEE Trans. Circuits Syst.}, vol. CAS-30,
  no.~3, pp. 148--159, Sep. 1983.

\bibitem{kaye1982}
R.~J. Kaye and F.~F. Wu, ``Dynamic security regions of power systems,''
  \emph{IEEE Trans. Circuits Syst.}, vol. CAS-29, no.~9, pp. 612--623, Mar.
  1982.

\bibitem{young2014}
W.~Young and N.~G. Leveson, ``An integrated approach to safety and security
  based on systems theory,'' \emph{Comm. ACM}, vol.~57, no.~2, pp. 31--35, Feb.
  2014.

\bibitem{pavella1977}
M.~Ribbens-Pavella, B.~Lemal, and W.~Pinard, ``On-line operation of {Liapunov}
  criterion for transient stability studies,'' \emph{IFAC Proceedings Volumes},
  vol.~10, no.~1, pp. 292--296, Feb. 1977.

\bibitem{fouad1981}
A.~A. Fouad, ``Transient stability margin as a tool for dynamic security
  assessment,'' Dept. of Electrical Engineering, Iowa State University of
  Science and Technology, Tech. Rep. {\#}EPRI-EL-1755, 1981.

\bibitem{pavella1982}
M.~Ribbens-Pavella, P.~G. Murthy, J.~L. Horward, and J.~L. Carpentier,
  ``Transient stability index for online stability assessment and contingency
  evaluation,'' \emph{Int. J. Electr. Power Energy Syst.}, vol.~4, no.~2, pp.
  91--99, Apr. 1982.

\bibitem{chiang2010}
H.-D. Chiang, \emph{Direct Methods For Stability Analysis of Electric Power
  Systems: Theoretical Foundation, BCU Methodologies, and Applications}.\hskip
  1em plus 0.5em minus 0.4em\relax New York: Wiley, 2010.

\bibitem{xue1989}
Y.~Xue, T.~V. Cutsem, and M.~Ribbens-Pavella, ``Extended equal area criterion
  justifications, generations, applications,'' \emph{IEEE Trans. Power Syst.},
  vol.~4, no.~1, pp. 44--52, Nov. 1989.

\bibitem{hiskens2000}
I.~A. Hiskens and M.~A. Pai, ``Trajectory sensitivity analysis of hybrid
  systems,'' \emph{IEEE Trans. Circuits Syst. I, Fundam. Theory}, vol.~47,
  no.~2, pp. 204--220, Feb. 2000.

\bibitem{saito1975}
O.~Saito, K.~Koizumi, M.~Udo, M.~Sato, H.~Mukae, and T.~Tsuji, ``Security
  monitoring systems including fast transient stability studies,'' \emph{IEEE
  Trans. Power App. Syst.}, vol. PAS-94, no.~5, pp. 1789--1805, Sep./Oct. 1975.

\bibitem{chu2004}
X.~Chu and Y.~Liu, ``Real-time transient stability prediction using incremental
  learning algorithm,'' in \emph{2004 IEEE Power Engineering Society General
  Meeting (PESGM)}, vol.~2, Jun. 2004, pp. 1565--1569.

\bibitem{shenoy2016}
N.~Shenoy and R.~Ramakumar, ``Power systems dynamic security assessment using
  fisher information metric,'' in \emph{2016 IEEE Power and Energy Society
  General Meeting (PESGM)}, Jul. 2016, pp. 1--5.

\bibitem{bemporad1999}
A.~Bemporad and M.~Morari, ``Control of systems integrating logic dynamics and
  constraints,'' \emph{Automatica}, vol.~35, no.~3, pp. 407--427, Mar. 1999.

\bibitem{henzinger1996}
T.~A. Henzinger, ``The theory of hybrid automata,'' in \emph{Proceedings of the
  7th Annual IEEE Symposium on Logic in Computer Science}, 1996, pp. 278--292.

\bibitem{tomlin2000}
C.~J. Tomlin, J.~Lygeros, and S.~S. Sastry, ``A game theoretic approach to
  controller design for hybrid systems,'' \emph{P. IEEE}, vol.~88, no.~7, pp.
  949--970, Jul. 2000.

\bibitem{lygeros2003}
J.~Lygeros, K.~H. Johansson, S.~N. Simic, J.~Zhang, and S.~S. Sastry,
  ``Dynamical properties of hybrid automata,'' \emph{IEEE Trans. Autom.
  Control}, vol.~48, no.~1, pp. 2--17, Jan. 2003.

\bibitem{lygeros2006}
J.~Lygeros, \emph{Lecture Notes on Hybrid Systems}.\hskip 1em plus 0.5em minus
  0.4em\relax ETH Zuich, 2006.

\bibitem{fourlas2004}
G.~K. Fourlas, K.~J. Kyriakopoulos, and C.~D. Vournas, ``Hybrid systems
  modeling for power systems,'' \emph{IEEE Circuits Syst. Mag.}, vol.~4, no.~3,
  pp. 16--23, Third Quarter 2004.

\bibitem{hikihara2005}
T.~Hikihara, ``Application of hybrid system theory to power system analysis
  ({I}) (in {Japanese}),'' in \emph{Annual Meeting Record IEE Japan}, vol.~6,
  Mar. 2005, p. 187.

\bibitem{susuki2009}
Y.~Susuki, Y.~Takatsuji, and T.~Hikihara, ``Hybrid model for cascading outage
  in a power system: A numerical study,'' \emph{IEICE Trans. Fundam. Electron.
  Commun. Comput. Sci.}, vol. E92-A, no.~3, pp. 871--879, Mar. 2009.

\bibitem{susuki2012}
Y.~Susuki, T.~J. Koo, H.~Ebina, T.~Yamazaki, T.~Ochi, T.~Uemura, and
  T.~Hikihara, ``A hybrid system approach to the analysis and design of power
  grid dynamic performance,'' \emph{P. IEEE}, vol. 100, no.~1, pp. 225--239,
  Jan. 2012.

\bibitem{susuki2015}
Y.~Susuki and T.~J. Koo, ``An application of {RRT} algorithm to reliability
  assessment of energy systems (in {Japanese}),'' IEICE, Tech. Rep.
  {\#}NLP2014-128, Jan. 2015.

\bibitem{wu2016}
Q.~Wu, Y.~Susuki, and T.~J. Koo, ``{RRT}-based computation for dynamic security
  analysis of power systems,'' \emph{IEICE Trans. Fundam. Electron. Commun.
  Comput. Sci.}, vol. E99-A, no.~2, pp. 491--501, Feb. 2016.

\bibitem{amato1996}
N.~M. Amato and Y.~Wu, ``A randomized roadmap method for path and manipulation
  planning,'' in \emph{Proceedings of the 1996 IEEE International Conference on
  Robotics and Automation}, 1996, pp. 113--120.

\bibitem{kavraki1996}
L.~E. Kavralu, P.~Svestka, J.-C. Latombe, and M.~H. Overmars, ``Probabilistic
  roadmaps for path planning in high-dimensional configuration spaces,''
  \emph{IEEE Trans. Robot. Autom.}, vol.~12, no.~4, pp. 566--580, Aug. 1996.

\bibitem{lavalle1998}
S.~M. Lavalle, ``Rapidly-exploring random trees: {A} new tool for path
  planning,'' Computer Science Dept., Iowa State University, Tech. Rep.
  {\#}98-11, Oct. 1998.

\bibitem{lavalle2006}
------, \emph{Planning Algorithms}.\hskip 1em plus 0.5em minus 0.4em\relax
  Cambridge: Cambridge University Press, 2006.

\bibitem{lavalle2001}
S.~M. Lavalle and J.~J. {Kuffner, Jr.}, ``Rapidly-exploring random trees:
  {P}rogress and prospects,'' in \emph{Algorithmic and Computational Robotics:
  New Directions}, B.~Donald, K.~Lynch, and D.~Rus, Eds.\hskip 1em plus 0.5em
  minus 0.4em\relax Wellesley, MA: A. K. Peters, 2001, pp. 293--308.

\bibitem{lavalle2000}
S.~M. Lavalle, ``Robot motion planning: A game-theoretic foundation,''
  \emph{Algorithmica}, vol.~26, no.~3, pp. 430--465, Apr. 2000.

\bibitem{lavalle1999}
S.~M. Lavalle and J.~J. {Kuffner, Jr.}, ``Randomized kinodynamic planning,'' in
  \emph{Proceedings of the 1999 IEEE International Conference on Robotics and
  Automation}, 1999, pp. 473--479.

\bibitem{lavalle2001a}
------, ``Randomized kinodynamic planning,'' \emph{Int. J. Rob. Res.}, vol.~20,
  no.~5, pp. 378--400, 2001.

\bibitem{branicky2003}
M.~S. Branicky, M.~M. Curtiss, J.~A. Levine, and S.~B. Morgan, ``{RRTs} for
  nonlinear, discrete, and hybrid planning and control,'' in \emph{Proceedings
  of the 42nd IEEE Conference on Decision and Control}, 2003, pp. 657--663.

\bibitem{bhatia2004}
A.~Bhatia and E.~Frazzoli, ``Incremental search methods for reachability
  analysis of continuous and hybrid systems,'' in \emph{Proceedings of the 7th
  International Workshop on Hybrid Systems: Computation and Control}, ser.
  Lecture Notes in Computer Science, R.~Alur and G.~J. Pappas, Eds., vol.
  2993.\hskip 1em plus 0.5em minus 0.4em\relax Berlin, Heidelberg: Springer
  Berlin Heidelberg, 2004, pp. 142--156.

\bibitem{dang2006}
T.~Dang and T.~Nahhal, ``Randomized simulation of hybrid systems for circuit
  validation,'' in \emph{Proceedings of the Forum on Specification and Design
  Languages}, Sep. 2006, pp. 9--15.

\bibitem{dang2009}
------, ``Coverage-guided test gerneration for continuous and hybrid systems,''
  \emph{Form Methods Syst. Des.}, vol.~34, pp. 183--213, 2009.

\bibitem{plaku2007}
E.~Plaku, L.~E. Kavraki, and M.~Y. Vardi, ``Hybrid systems: {F}rom verification
  to falsification,'' in \emph{Proceedings of 19th International Conference on
  Computer Aided Verification}, ser. Lecture Notes in Computer Science, W.~Damm
  and H.~Hermanns, Eds., vol. 4590.\hskip 1em plus 0.5em minus 0.4em\relax
  Springer, Jul. 2007, pp. 463--476.

\bibitem{plaku2009}
------, ``Hybrid systems: {F}rom verification to falsification by combining
  motion planning and discrete search,'' \emph{Form. Methods Syst. Des.},
  vol.~34, no.~2, pp. 157--182, Apr. 2009.

\bibitem{xu2010}
J.~J. Xu, T.~J. Koo, and Z.~X. Li, ``Sampling-based finger gaits planning for
  multifingered robotic hand,'' \emph{Auton. Robot.}, vol.~28, no.~4, pp.
  385--402, May 2010.

\bibitem{brenan1996}
K.~E. Brenan, S.~L. Campbell, and L.~R. Petzold, \emph{Numerical Solution of
  Initial-Value Problems in Differential-Algebraic Equations}.\hskip 1em plus
  0.5em minus 0.4em\relax Philadelphia: Society for Industrial and Applied
  Mathematics, 1996.

\bibitem{chow1992}
J.~H. Chow and K.~W. Cheung, ``A toolbox for power system dynamics and control
  engineering education and research,'' \emph{IEEE Trans. Power Syst.}, vol.~7,
  no.~4, pp. 1559--1564, Nov. 1992.

\bibitem{milano2005}
F.~Milano, ``An open source power system analysis toolbox,'' \emph{IEEE Trans.
  Power Syst.}, vol.~20, no.~3, pp. 1199--1206, Aug. 2005.

\bibitem{abdu2017}
A.~Humayed, J.~Lin, F.~Li, and B.~Luo, ``Cyber-physical systems security {-- A}
  survey,'' \emph{IEEE Internet Things J.}, vol.~4, no.~6, pp. 1802--1831, Dec.
  2017.

\bibitem{cintuglu2017}
M.~H. Cintuglu, O.~A. Mohammed, K.~Akkaya, and A.~S. Uluagac, ``A survey on
  smart grid cyber-physical system testbeds,'' \emph{IEEE Commun. Surveys
  Tuts.}, vol.~19, no.~1, pp. 446--464, Feb. 2017.

\bibitem{morison2004}
K.~Morison, L.~Wang, and P.~Kundur, ``Power system security assessment,''
  \emph{IEEE Power and Energy Magazine}, vol.~2, no.~5, pp. 30--39, Sep. 2004.

\bibitem{kuffner2000}
J.~James J.~Kuffner and S.~M. LaValle, ``Rrt-connect: An efficient approach to
  single-query path planning,'' in \emph{Proceedings of the IEEE International
  Conference on Robotics and Automation}, vol.~2, 2000, pp. 995--1001.

\bibitem{karaman2010}
S.~Karaman and E.~Frazzoli, ``Incremental sampling-based algorithms for optimal
  motion planning,'' in \emph{Proceedings of the Robotics: Science and
  Systems}, Jun. 2010, pp. 1--8.

\bibitem{perez2012}
A.~Perez, R.~{Platt Jr.}, G.~Konidaris, L.~Kaelbling, and T.~Lozano-Perez,
  ``{LQR-RRT$^*$}: Optimal sampling-based motion planning with automatically
  derived extension heuristics,'' in \emph{Proceedings of the IEEE
  International Conference Robotics and Automtion}, May 2012, pp. 2537--2542.

\bibitem{sucan2012}
I.~A. {\c{S}}ucan, M.~Moll, and L.~E. Kavraki, ``The {O}pen {M}otion {P}lanning
  {L}ibrary,'' \emph{{IEEE} Robotics \& Automation Magazine}, vol.~19, no.~4,
  pp. 72--82, Dec. 2012.

\bibitem{sauer1998}
P.~W. Sauer and M.~A. Pai, \emph{Power System Dynamics and Stability}.\hskip
  1em plus 0.5em minus 0.4em\relax Upper Saddle River, New Jersey 07458:
  Prentice Hall, 1998.

\bibitem{bagheri2013}
A.~Bagheri, ``Optimal location and signal selection of {UPFC} device for
  damping oscillation,'' Master's thesis, Najaf Abad Branch, Islamic Azad
  University, 2013.

\bibitem{machowski1997}
J.~Machowski, J.~W. Bialek, and J.~R. Bumby, \emph{Power System Dynamics:
  Stability and Control}.\hskip 1em plus 0.5em minus 0.4em\relax John Wiley
  {\&} Sons, 1997.

\bibitem{susuki2011}
Y.~Susuki, I.~Mezi{\'{c}}, and T.~Hikihara, ``Coherent swing instability of
  power grids,'' \emph{J. Nonlinear Sci.}, vol.~21, no.~3, pp. 403--439, Feb.
  2011.

\bibitem{kundur1994}
P.~Kundur, \emph{Power System Stability and Control}.\hskip 1em plus 0.5em
  minus 0.4em\relax McGraw-hill, 1994.

\bibitem{yang2013}
Y.~Yang, ``Hardware acceleration of power system simulation,'' Master's thesis,
  Imperial College London, 2013.

\end{thebibliography}


%
%
%
%
%
%

\end{document}